\documentclass[12pt]{article}

\usepackage[
top    = 1in,
bottom = 1in,
left   = 1in,
right  = 1in]{geometry}

  \usepackage[dvipsnames]{xcolor}
  \usepackage{adjustbox, setspace}
  \usepackage{caption,subcaption,wrapfig}
\usepackage[normalem]{ulem} 
 \useunder{\uline}{\ul}{} 
 \usepackage{amsmath,multirow, bm, titling}
 
 \usepackage{tikz}
 \usepackage{pgf}
\usetikzlibrary{shapes.geometric, arrows, positioning}

\usepackage[numbers]{natbib} 
 \usepackage{fancyhdr}
 \newtheorem{theorem}{Theorem}

\newtheorem{lemma}[theorem]{Lemma}

\newtheorem{definition}[theorem]{Definition}

\newtheorem{algorithm}[theorem]{Algorithm}

 \newcommand\ben{\begin{enumerate}}
 \newcommand\een{\end{enumerate}}
  \newcommand\beq{\begin{eqnarray}}
 \newcommand\eeq{\end{eqnarray}}
   \newcommand\beqa{\begin{eqnarray*}}
 \newcommand\eeqa{\end{eqnarray*}}
   \newcommand\bal{\begin{aligned}}
 \newcommand\eal{\end{aligned}}
  \newcommand\bi{\begin{itemize}}
 \newcommand\ei{\end{itemize}}

\newcommand{\0}{\bm{\0}}

\begin{document}

\title{Predictive case control designs for modification learning}
\author{W. Katherine Tan, Patrick J. Heagerty}
\date{}
\begin{titlingpage}
    \maketitle
    \begin{abstract}
    \noindent Prediction models for clinical outcomes, originally developed using a source dataset, may additionally be applied to a new cohort. Prior to accurate application, the source model needs to be updated through external validation procedures such as modification learning. Model modification learning involves the dual goals of recalibrating overall predictions as well as revising individual feature effects, and generally requires the collection of an adequate sample of true outcome labels from the new setting. Unfortunately, outcome label collection is frequently an expensive and time-consuming process, as it involves abstraction by human clinical experts. To reduce the abstraction burden for such new data collection, we propose a class of designs based on original model scores and their associated outcome predictions. We provide mathematical justification that the general predictive score sampling class results in valid samples for analysis. Then, we focus attention specifically on a stratified sampling procedure that we call predictive case control (PCC) sampling, which allows the dual modification learning goals to be achieved at a smaller sample size compared to simple random sampling (SRS). PCC sampling intentionally over-represents subjects with informative scores, where we suggest using the D-optimality and Binary Entropy information functions to summarize sample information. For design evaluation within the PCC class, we provide a computational framework to estimate and visualize empirical response surfaces of the proposed information functions. We demonstrate the benefit of using PCC designs for modification learning, relative to SRS, through Monte Carlo simulation. Finally, using radiology report data from the Lumbar Imaging with Reporting of Epidemiology (LIRE) study, we illustrate the application of PCC for new outcome label abstraction and subsequent modification learning across imaging modalities.
    \end{abstract}
\end{titlingpage}

\section{Introduction}
\label{sec:pcc_intro}
\noindent Clinical prediction models are often developed to help determine a specific diagnosis, to inform prognosis, and to define subgroups relevant for various clinical outcomes. Often, a clinical prediction model is developed based on data drawn from a source cohort, for example a particular hospital or research network. To facilitate sharing of learned patterns from established clinical data sources, a developed model may also be modified and applied to new settings, for example different health systems, age groups, or time frames. Before adopting model predictions in the new setting, it is important to ask whether model predictions provide an accurate representation of the true risks in the target population. For example, is there sufficient agreement between observed outcomes and model predictions? Can the predictions sufficiently distinguish between cases and controls? In an ideal situation, a model would be perfectly generalizable to the new setting without any modification. Unfortunately, a model can be valid in the source setting yet invalid when applied to new cohorts, due to reasons such as case mix differences, model over-fitting, or true differences between the populations \cite{steyerberg2008clinical}. Therefore, the developed model needs to be assessed for validity in the new setting, and modified appropriately \cite{steyerberg2008clinical, steyerberg2014towards}.\\ 

\noindent Data driven model updating or modification requires the collection of an adequate sample from the new setting. Outcome labels are often labor intensive to collect, as they require abstraction from the medical record, yet substantial sample sizes are necessary for modification learning studies. For example, systematic recalibration adjustment requires $312$ observed events if predictions were 25\% too extreme on the odds scale \cite{vergouwe2005substantial}. For rare outcomes, such substantial sample size requirements may be difficult to obtain without larger or targeted samples. In this paper, we motivate and describe a sampling strategy targeted for the modification learning problem. The proposed approach leverages the original model scores in order to design efficient data collection. Our proposed sampling design is motivated by strategies from experimental design and machine-learning, and may reduce the sample size requirements for model modification learning and assessment.

\section{Background}
\label{sec:pcc_background}
\subsection{Modification learning of clinical prediction models}

\noindent When applying a previously developed model to a new setting, predictions based on the original model may be invalid, due to reasons such as differences in case mix or differences in regression coefficients \cite{steyerberg2008clinical}. Case mix is when the distribution of features or outcomes are different between the source and new setting. Differences in regression coefficients may occur from model over-fitting on small development datasets or from truly different populations due to different cohort selection criteria. Updating or modification learning of a prediction model to new settings should involve empirical procedures such as model recalibration, revision, and extension \cite{steyerberg2008clinical, steyerberg2014towards}.\\

\noindent Model recalibration involves the systematic adjustment of potentially inaccurate predictions. However, since unnecessary recalibration may introduce variation for model application, whether the original model is sufficiently calibrated in the new setting may be first formally tested. For example, testing can be based on the Hosmer-Lemeshow test \cite{hosmer1997comparison}, where estimated and observed predictions are compared across strata of grouped averages. Since the Hosmer-Lemeshow test involves variable or arbitrary binning choices, the logistic recalibration \cite{cox1958two} test is potentially more powerful, where specific deviations from the null hypothesis indicate if recalibration is required. If model recalibration is deemed necessary, the Cox logistic recalibration model \cite{cox1958two} adjusts overall mean predictions as well as systematically overestimated and underestimated predictions. To increase flexibility, the linear-logistic assumption may be relaxed by modeling additional shape parameters in the sigmoid function \cite{stukel1988generalized} or non-linear functions of the predicted scores \cite{dalton2013flexible}.\\

\noindent Model revision is a step beyond recalibration that involves re-estimating individual feature effects instead of overall systematic adjustments. Previously estimated coefficients can be refitted in a stepwise manner \cite{steyerberg2004validation} or through shrinkage and penalized models for larger numbers of coefficients \cite{moons2004penalized}. In addition, the original model may be extended to include additional features. Model revision can be done without recalibration, where the model is completely re-fitted to the new data \cite{steyerberg2004validation}. Since complete revision may potentially be unstable and ignores past knowledge, a more strategic approach combines both recalibration and revision in the same model. If data is collected in batches from the new setting, models may also be continuously modified, where model recalibration is suggested to take precedence over model revision and/or extension \cite{steyerberg2008clinical}.

\subsection{Information criteria for modification learning}

\noindent For modification learning, an adequate sample of actual outcome labels needs to be collected from the new setting. Due to the potential substantial sample size requirements \cite{vergouwe2005substantial} under simple random sampling (SRS), an alternative is to consider sampling designs specifically targeted towards the modification learning problem. To evaluate the ``information'' of samples for modification learning, here we review relevant statistical criteria from the related fields of information theory and optimal experimental design.

\subsubsection*{Information theoretic criteria}

\begin{figure}[!htbp]
\centering 
\caption{Illustration of Binary Entropy as a function of sample outcome prevalence.}
\label{pcc_fig:binary_entropy}
\includegraphics[width=2.75in,height=1.5in]{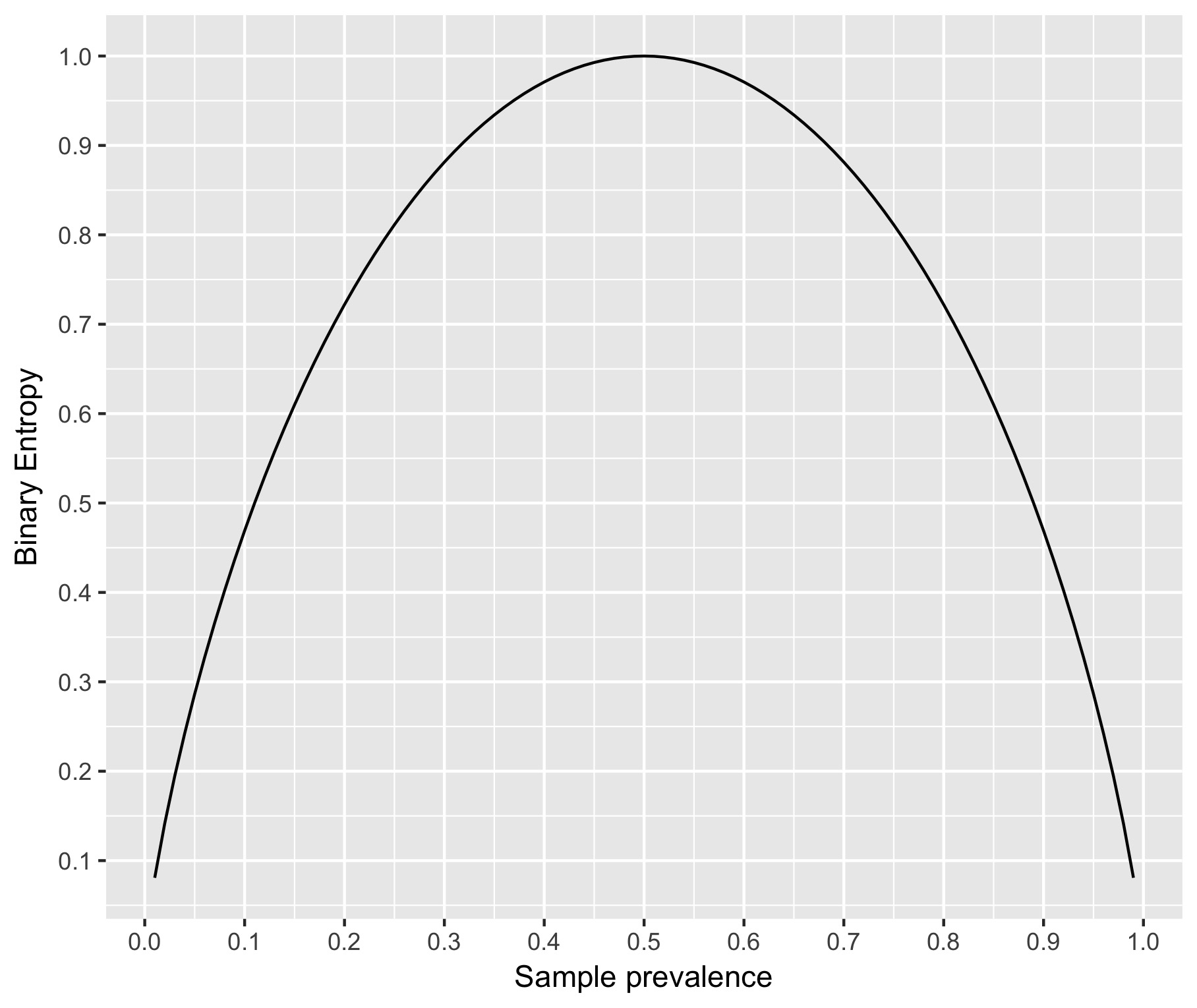} 
\end{figure}

\noindent In information theory, objective criteria to measure information in a given sample is generally based on maximizing uncertainty or information. For example, Binary Entropy \cite{mackay2003information} measures the amount of uncertainty for a binary variable (see Figure \ref{pcc_fig:binary_entropy} for an illustration). If subjects in the samples are either all cases or all controls, there is no variation in outcome labels and thus no information for learning - the Binary Entropy is at its minimum of $0$. On the other hand, if the sample contains approximately equal cases and controls, then information is maximized and Binary Entropy attains its maximum of $1$. Maximizing sample Binary Entropy may help classification, since samples with approximately outcome class balance are generally accepted to improve classifier learning \cite{weiss2001effect,batista2004study,xue2015does}. Another approach of quantifying uncertainty is based on distance to the decision boundary. For example, subjects with initial predicted probabilities close to the cut-off for binary decisions may be considered to have ``uncertain'' predictions, therefore collecting their true outcome statuses may provide more information.

\subsubsection*{Statistical information criteria}

\noindent The field of optimal experimental designs offers theoretical guidance for study design when predictors can be chosen, outcomes need to be collected, and practical constraints limit the number of experimental runs. For a given statistical model and associated parameters, information for parameter estimation can be maximized using one-dimensional summaries of the information matrix (summarized in Table \ref{pcc_tb:optimality_criteria}). Many such criteria are considered ``true'' information functions, having the monotonicity, concavity, and homogeneity properties for comparing matrices ordered in the Loewner sense \cite{pukelsheim1993optimal}. This means that if the difference of two information matrices is positive semi-definite, then associated one-dimensional summaries are similarly ordered. For a fixed sample size, subjects can be allocated based on pre-specified criteria, where the unique predictor values are called ``design points'', and the proportions of subjects at each design point are called ``design weights''. Optimal statistical designs provide convenient interpretations, for example a design targeting maximizing D-optimality is one that minimizes the joint confidence region of model parameters \cite{eriksson2000design}.\\

\noindent Even though optimal design theory provides an intuitive framework for study planning, finding the optimal design in practice can be a very challenging problem. Early work discussed `'`exact'' optimal designs, which is based on identifying combinations of subjects that result in maximum sample information \cite{atkinson1982developments}. Alternatively, approximately optimal designs places a probability distribution on the design space, thus allowing for non-discrete design points \cite{kiefer1959optimum}. For example, in stratified designs, the design space is decomposed into non-overlapping partitions and non-zero weights are placed on each stratum. Even for approximately optimal designs, there is not a single unique ``best'' design, motivating using Monte Carlo and approximation-based algorithms \cite{atkinson2015designs}. In addition, within a class of designs, pre-specified statistical criteria can also be used to evaluate competing configurations, and higher information designs are selected for data collection.\\

\noindent For logistic regression, since the information matrix depends on true parameters, optimal designs need to be computed on true or assumed model parameters and therefore are only locally optimal \cite{khuri2006design, atkinson2015designs}. For example, in estimating a one-parameter logistic regression model, the locally D-optimal design places equal weight on two symmetric points, but design point placement depends on true parameter values. To account for a range of possible parameter values, approaches such as minimax, sequential, and Bayesian methods can be used \cite{chipman1996d,khuri2006design, atkinson2015designs}.\\

\begin{table}[h!]
\centering
\caption{Selected common statistical optimality criteria, mathematical formulae, and interpretation. $\mathbf{I_m}$ is the Fisher's information matrix and $\mathbf{H}$ is the hat/projection matrix, where for logistic regression $\mathbf{I_m} = \mathbf{X}^T \mathbf{W}\mathbf{X}$ and $\mathbf{H} = \mathbf{W}^{1/2} \mathbf{X} \mathbf{I_m}^{-1} \mathbf{X}^T \mathbf{W}^{1/2}$, $w_{ii} = p_i(1-p_i)$.}
\label{pcc_tb:optimality_criteria}
\begin{tabular}{|c|c|c|}
\hline
\textbf{Criterion} & \textbf{Mathematical Formula} & \textbf{Interpretation of using criteria} \\ \hline
A-optimality & $\underset{\zeta}{argmin} \: trace(\mathbf{I_m}^{-1})$ & Minimizes average variance of parameters. \\ \hline
C-optimality & $\underset{\zeta}{argmin} \: trace(c^T \mathbf{I_m}^{-1} c)$ & Minimizes variance of a best linear unbiased estimator. \\ \hline
D-optimality & $\underset{\zeta}{argmax} \: det(\mathbf{I_m})$ & Minimizes volume of parameter joint confidence region. \\ \hline
G-optimality & $\underset{\zeta}{argmin} \: \underset{\zeta}{argmax}\: Diag(\mathbf{H})$ & Minimizes the maximum variance of predicted values. \\ \hline
\end{tabular}
\end{table}
    
\subsection{Related problems and research gap}

\noindent Modification learning of clinical prediction models is related to transfer learning and active learning from machine-learning; here we summarize these related frameworks to motivate our proposed statistical design approach. 

\subsubsection*{Transfer learning}

\noindent Transfer learning involves the partial modification of an original model to the new setting using source data. Transfer learning strategies may be categorized as either transductive or inductive \cite{pan2010survey}, both of which have connections to modification learning. For transductive learning, similar to the case-mix assumption, feature distributions $f(x)$ between source and new scenarios are assumed to be different. Transductive learning approaches generally assume that no outcome labels are available from the new setting, therefore distributional differences are addressed by fitting a weighted model to source data. On the other hand, inductive learning assumes that conditional outcome distributions $f(y|x)$ are different between source and target, which is a generalization of the different coefficient assumption. For inductive learning, some outcome labels from the new setting are available, but the sample size is too small to generate reliable model predictions. Instead, assuming certain shared parameters, the existing model only needs to be partially modified for transfer to the new setting. In the biomedical setting, transfer learning has been successfully applied to skin cancer classification \cite{esteva2017dermatologist}, but required over 1 million outcome labels for the original model and over 100,000 labels for the partial modification to the new setting.

\subsubsection*{Active learning}

\noindent Another related problem is active learning, which is the sequential modification of an original model through querying new data. Several of the previously described information theoretic and statistical information criteria have been incorporated into active learning strategies. For example, pool-based active learning sequentially selects the most informative data points for outcome labeling from a pre-specified cohort ``pool''. In pool-based active learning, sample selection may be based on heuristically the most uncertain predictions, for example using the Binary Entropy criterion \cite{settles2012active}, or based on variance reduction, for example using the A-optimality criterion \cite{schein2007active}. Curiously, while some researchers advocate designs based on statistical information criteria \cite{schein2007active}, others have demonstrated that heuristic information criteria such as Binary Entropy perform well in empirical experiments \cite{yang2018benchmark}.\\

\noindent Active learning has been applied to machine-learning phenotyping models \cite{chen2013applying}, and for classification of cancer from radiology reports \cite{nguyen2014supervised}. Such applications have noted some sample size savings by using active learning over ``passive'' learning using a random or convenient sample. However, active learning algorithms have been demonstrated to perform even worse than SRS in some settings \cite{schein2007active}. A common criticism is that sequential myopic selection of subjects for outcome labeling without theoretical foundations may introduce unanticipated variation \cite{settles2012active}. Furthermore, implementing active learning often requires extensive engineering infrastructure and expertise. In fact, more than half of surveyed applied machine-learning researchers revealed hesitated to use active learning strategies for annotation or abstraction tasks \cite{tomanek2009web}. A concern is due to the unknown introduced bias from resulting samples: even if sampling with active learning may improve prediction accuracy, it is often difficult to use such samples for statistical analysis such as formal inference and evaluation.

\subsubsection*{Research gap and contribution}

\noindent In transfer learning, an existing model is partially modified, but new outcome labels are not necessarily collected. On the other hand, the active learning framework involves sequentially procuring of additional outcome labels based on pre-specified statistical criteria, so to achieve learning goals in a resource efficient manner. However, as most active learning algorithms frame data collection as a learning rather than traditional design problem, statistical inference is usually not a goal and therefore bias in resulting samples may not be well-characterized. In the clinical prediction setting, both sample generalizability and resource efficiency are important considerations. Furthermore, modification learning of clinical prediction models are usually based on existing starting points, for example previously published model coefficients and scoring rules. Therefore, this motivates a design-based framework for new outcome label collection and subsequent modification learning, particularly for practical settings where personnel and monetary resources constraint abstraction label sample sizes.

\section{Methods}
\label{sec:pcc_methods}
\subsection{Statistical notation and motivation}
\label{pcc_sec:methods_1notation}

\noindent Denote the source cohort as $\mathcal{D}^0$ and the new cohort as $\mathcal{D}^1$. For subject $i$ denote features as $\tilde{X}^T_i \in \mathcal{R}^p$ and binary outcome labels as $Y_i$. We assume that the original clinical prediction model denoted $\hat{h}(.)$ was developed using a sample drawn from $\mathcal{D}^0$. Applying $\hat{h}(.)$ to the data in $\mathcal{D}^1$ will generate original model prediction scores in the new data denoted as 

\beq\bal 
S_i &= \hat{h}(\tilde{X}^T_i).
\label{pcc_eq:score_def}
\eal\eeq 

\noindent The original model $\hat{h}(.)$ may need to be modified before ultimate reliable application to the new setting. Predictions may be recalibrated and model parameters potentially revised. We consider simultaneous recalibration and revision models \cite{steyerberg2004validation} of the form

\beq\bal 
logit(E[Y_i|S_i,\tilde{X}^T_i]) &= \alpha_0 + \alpha_1 S_i + \tilde{X}^T_i \tilde{\gamma}.
\label{pcc_eq:model}
\eal\eeq 

\noindent In \eqref{pcc_eq:model}, $\alpha_0$, $\alpha_1$, and $\tilde{\gamma}$ are the model modification learning parameters to be estimated. The recalibration parameters $\alpha_0$ and $\alpha_1$ re-adjust any systematic mis-estimation in predicted scores, while the revision parameters $\tilde{\gamma}$ identify any new predictive features or refinement of coefficients for original features. Assuming sparsity, \eqref{pcc_eq:model} can be modeled using the Lasso procedure \cite{tibshirani1996regression},

\beq\bal 
(\hat{\alpha}_0,\hat{\alpha}_1,\hat{\tilde{\gamma}}) = \underset{\alpha_0,\alpha_1,\tilde{\gamma}}{argmin}
\left\{ \sum\limits_{i=1}^{n} -Y_i(\alpha_0 + \alpha_1 S_i + \tilde{X}^T_i \tilde{\gamma} + 
\log(1 + \exp(\alpha_0 + \alpha_1 S_i + \tilde{X}^T_i \tilde{\gamma} )) + 
\lambda||\tilde{\gamma}||_1 \right\},
\label{pcc_eq:lasso_procedure}
\eal\eeq 

\noindent which uses penalization for estimation. Note that in \eqref{pcc_eq:lasso_procedure}, only revision parameters $\tilde{\gamma}$ are penalized, so that original scores $S$ are retained in the model. To fit model \eqref{pcc_eq:model} requires the collection of a sample drawn from the new setting $\mathcal{D}^1$, where features $\tilde{X}^T_i$, scores $S_i$, and outcome labels $Y_i$ are available for all subjects. We assume that $\tilde{X}^T_i$ is easily available, $S_i$ can be generated using \eqref{pcc_eq:score_def}, but outcome labels require expensive and time-consuming abstraction. Motivated by information theoretic and statistical information criteria, we describe a sampling design framework based on scores $S$, so to select subjects from the new setting for outcome label abstraction and subsequent model modification learning.
\subsection{Predictive Case Control (PCC) designs}
\label{pcc_sec:methods_2design}

\noindent For outcome label collection from the new setting, we define sampling strategies based on original model scores $S$ as the predictive score sampling class (Definition \ref{pcc_def:design_class}). The name ``predictive'' is due to that, arguably among all observed variables, scores $S$ are most predictive of unobserved true outcomes $Y$. Among possible ways to sample based on $S$, note that subjects with higher scores tend to be ``cases" ($Y=1$), while those with lower scores tend to be ``controls'' ($Y=0$).

\begin{definition} Predictive score sampling class.\\
For new outcome label collection, denote the class where sampling is based only on original model scores $S \in [0,1]$ as the predictive score sampling class.
\label{pcc_def:design_class}
\end{definition}

\noindent Among the class of predictive score sampling deigns, we define a stratified sampling procedure as the Predictive Case Control (PCC; Definition \ref{pcc_def:pcc}), which are indexed by design configurations defined with score cut-off $k$ and stratum weights $w = P(S>k|\text{sampled})$. For an interpretation of PCC, assume that the score distribution $f(S)$ is a mixture of marginal outcome case/control proportions $P(Y=y)$ and conditional score distributions $f(S|Y=y)$. Then, PCC design configurations $w$ and $k$ can be interpreted as intentionally altering $P(Y=y)$ and $f(S|Y=y)$ in resulting samples, respectively.

\begin{definition} Predictive case control (PCC) design.\\
The predictive case control design is a stratified sampling procedure based on original model scores $S \in [0,1]$, where for a fixed abstraction sample size $n$ and selected design configurations defined with score cut-off $k$ and stratum weights $w = P(S>k|\text{sampled})$, the procedure is

\beqa  
\text{Select:}
\begin{cases}
n \times w \:\: \text{subjects} & \text{from those with scores } \:\: S > k \\
n \times  (1-w) \:\: \text{subjects} & \text{from those with scores } \:\: S \leq k.
\end{cases}
\eeqa 
\label{pcc_def:pcc}
\end{definition}

\begin{figure}[!h]
\caption{Left: Score distribution by outcome classes of a simulated cohort with $N=10,000$; Right: Score distribution by outcome classes of a sample ($n=300$) drawn from the cohort using a Predictive Case Control design with configurations $k=2.5$, $w=0.50$.}
\label{pcc_fig:pcc_illus}
\centering
\includegraphics[width=3in,height=2in]{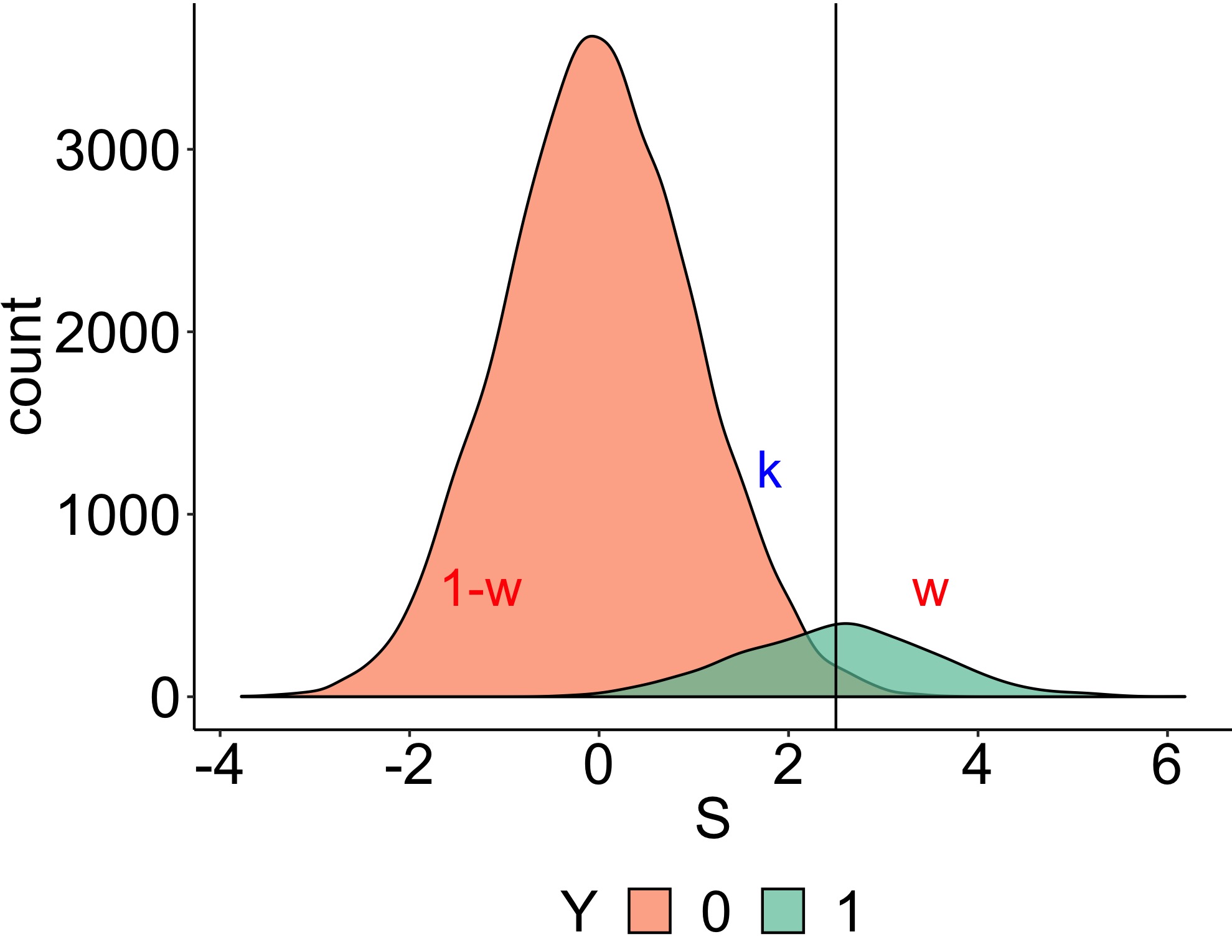}
\includegraphics[width=3in,height=2in]{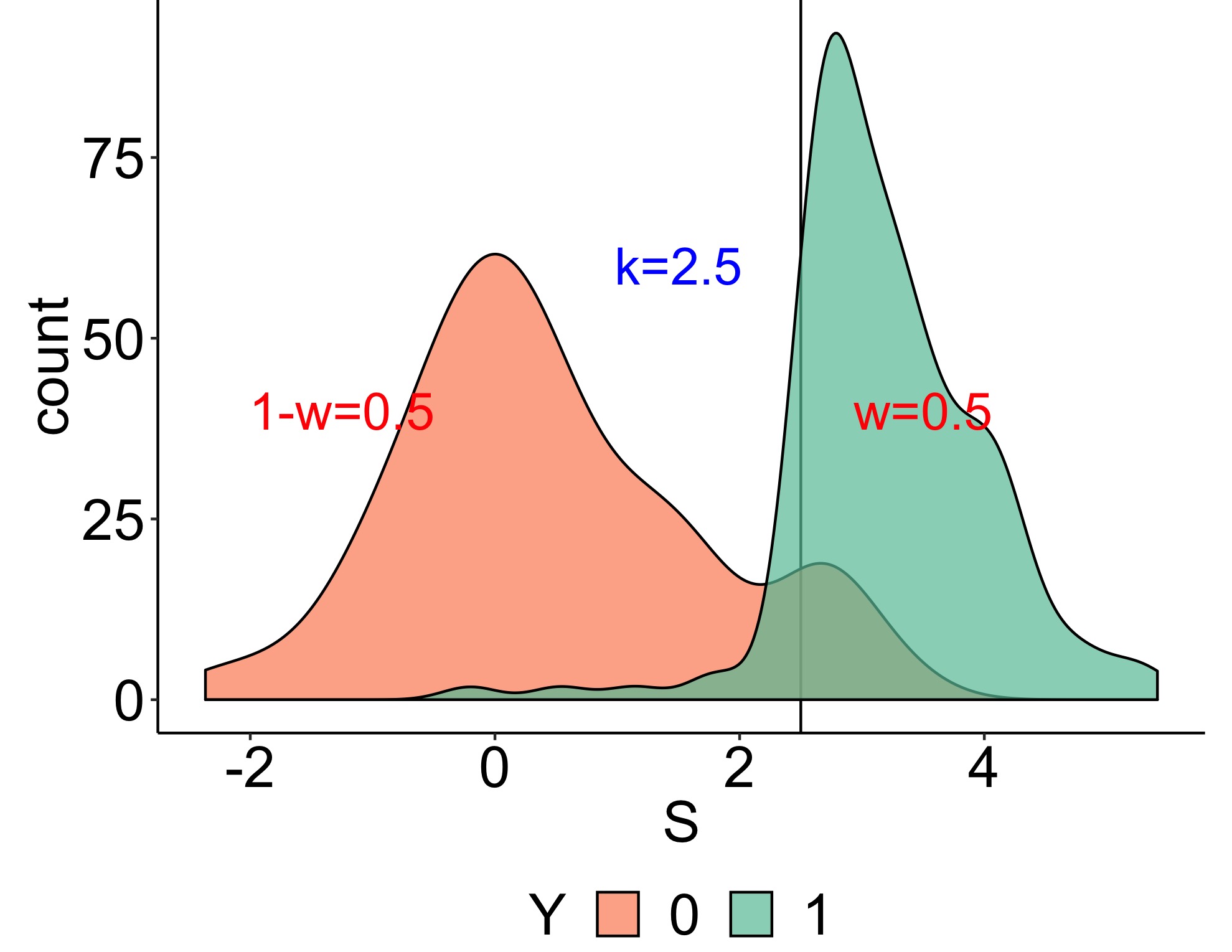}
\end{figure}

\noindent Figure \ref{pcc_fig:pcc_illus} illustrates the effect of using PCC design on induced sample score distribution, where sample scores may have different marginal outcome prevalence as well as conditional score distributions compared to that in the cohort. The intended benefit of using PCC is to reduce sample size requirements for modification learning and evaluation, through selecting subjects with ``high information'' for modeling goals. In particular, information functions of scores motivated by the dual modification learning goals of recalibration and revision may be used to select appropriate design configurations.

\subsubsection*{Recalibration goals: Statistical power and D-optimality}

\noindent For recalibration, since unnecessary readjustments may introduce additional variation, a first step is to test whether model recalibration is required. From the modification learning model \eqref{pcc_eq:model}, recalibration parameters $\alpha_0$ and $\alpha_1$ that deviate from $0$ and $1$ indicate potential mis-calibration in the new setting. For such recalibration testing, which may be based on Likelihood Ratio Tests in Table \ref{pcc_tb:recalibration_tests}, we consider maximizing statistical power as a relevant modeling goal.\\

\begin{table}[!h]
\centering
\caption{Model recalibration hypothesis tests. LRT = Likelihood Ratio Test. $\alpha_0,\alpha_1$ respectively indicate the recalibration intercept and slope.}
\begin{tabular}{p{5.5cm}p{2.9cm}p{3cm}p{4cm}}
 & Recalibration intercept & Recalibration slope & Logistic recalibration \\ \hline 
Null hypothesis, $H_0$ & $\alpha_0|(\alpha_1=1) = 0$ & $\alpha_1 = 1$ & $(\alpha_0, \alpha_1) = (0,1)$ \\ 
Alternative hypothesis, $H_A$ & $\alpha_0|(\alpha_1=1) \neq 0$ & $\alpha_1 \neq 1$ & $(\alpha_0, \alpha_1) \neq (0,1)$ \\ 
Degrees of freedom using LRT & 1 & 1 & 2 \\ \hline 
\end{tabular}
\label{pcc_tb:recalibration_tests}
\end{table}

\noindent Maximizing the local power of likelihood ratio statistical tests has been noted to be statistically equivalent to maximizing the determinant of the information matrix under the null \cite{wald1943tests}. Therefore, under the assumption that 

\beq\bal 
\alpha_0 = 0, \: \alpha_1 = 1, \: \tilde{\gamma} = \tilde{0},
\label{pcc_eq:null}
\eal\eeq 

\noindent we propose using the D-optimality criterion \cite{kiefer1959optimum} as a function to summarize information in sample scores. Define the score information function based on D-optimality criterion, $\phi^D(S)$, as

\beq\bal 
\phi^D(S) &:= \log \left(det \left( \mathbf{I}_m(S)\right)\right).
\label{pcc_eq:d_optimality}
\eal\eeq 

\noindent where in \eqref{pcc_eq:d_optimality}, $\mathbf{I}_{m}(S) = \frac{1}{n} \sum\limits_{i=1}^{n} \mathbf{I}_{mi}(S)$ is the sample partial information matrix for recalibration parameters ($\alpha_0, \alpha_1$), where for $p_i = \text{expit}(S_i)$,

\beqa\bal 
\mathbf{I}_{mi}(S) &= \begin{bmatrix}
p_i(1-p_i)  & S_i p_i(1-p_i)  \\
S_i p_i(1-p_i) & S^2_i p_i(1-p_i)
\end{bmatrix}
\eal\eeqa 

\noindent Note that $\phi^D(S)$ is always defined, as $\mathbf{I}_m(S)$ is by construction positive semi-definite. Designs with higher values of $\phi^D(S)$ may be interpreted as having lower generalized variances and smaller joint confidence regions of recalibration parameter estimates \cite{eriksson2000design}.

\subsubsection*{Revision goals: Support recovery and Binary Entropy}

\noindent Also using the modification learning model \eqref{pcc_eq:model}, revision parameters $\tilde{\gamma}$ that are non-zero indicate that model revision is necessary. Assuming sparsity for the revision parameters, an evaluation criterion for model revision performance can be based on the support recovery of truly predictive features. Denote the true revision parameters $\gamma_j^*$, $j = 1, \hdots, p$, where each $\gamma_j^*$ can be either truly predictive ($\gamma_j^* \neq 0$) or non-predictive ($\gamma_j^* = 0$), belonging to one of the mutually exclusive sets $\boldsymbol{\gamma_{S}^*}$ or $\boldsymbol{\gamma_{S^c}^*}$, defined as

\beq\bal 
\boldsymbol{\gamma_{S}^*} &= \{\gamma_j^*: \gamma_j^* \neq 0 \} \\
\boldsymbol{\gamma_{S^c}}^*&= \{\gamma_j^*: \gamma_j^* = 0 \}.
\label{pcc_eq:true_revision_param}
\eal\eeq 

\noindent Based on fitting Lasso procedure \eqref{pcc_eq:lasso_procedure}, denote the estimated revision parameters as $\hat{\gamma}_j$, $j=1, \hdots, p$, where each estimate $\hat{\gamma}_j$ is either predictive or non-predictive, with corresponding sets defined as

\beq\bal 
\boldsymbol{\hat{\gamma}_S} &= \{\hat{\gamma}_j: \hat{\gamma}_j \neq 0 \} \\
\boldsymbol{\hat{\gamma}_{S^c}} &= \{\hat{\gamma}_j: \hat{\gamma}_j = 0 \} 
\label{pcc_eq:estimated_revision_param}.
\eal\eeq 

\noindent Then, a measure for how well $\hat{\gamma}_j$ estimates true $\gamma_j^*$ can be represented using the False Discovery Rate (FDR) and False Exclusion Rate (FER) measures, where:

\beq\bal 
FDR &= 1-P(\boldsymbol{\hat{\gamma}_S} \in \boldsymbol{\gamma_{S}^*} ) \\
FER &= 1-P(\boldsymbol{\hat{\gamma}_{S^c}} \in \boldsymbol{\gamma_{S^c}^*}).
\label{pcc_eq:fdr_fer}
\eal\eeq  

\noindent In \eqref{pcc_eq:fdr_fer}, FDR can be interpreted as the false positive rate (Type I error), and FER the false negative rate (Type II error), in identifying the set of truly predictive features for model revision. Perfect support recovery is when both FDR and FER are zero. Under certain assumptions on the data generating mechanism\footnote{assumptions include that model dimensionality $p$ and sparsity ratio $f=\dfrac{k}{p}$, $k = |\boldsymbol{\gamma_{S}^*}|$ do not vary with sample size}, support recovery errors tend to decrease with sample size.\\

\noindent We take the view of improving support recovery for a fixed sample size. Note that the modification learning model \eqref{pcc_eq:model} may be framed as a classification model, where a key factor that affects classifier ``learning'' is outcome class balance in the development sample. Much research on outcome class balance on model prediction accuracy have focused on metrics such as the proportion correct (simple accuracy) or the Area Under the Receiving Operator Characteristic curve (AUC) \cite{weiss2001effect,batista2004study,xue2015does}. However, outcome class balance may additionally affect support recovery accuracy measures. Therefore, we propose using the Binary Entropy criterion as another summary of information based on scores. Define the information function based on Binary Entropy, $\phi^B(S)$, as 

\beq\bal 
\phi^B(S) &= -\bar{p}(S) log_2(\bar{p}(S) ) - (1-\bar{p}(S))log_2(1-\bar{p}(S)),
\label{pcc_eq:binary_entropy}
\eal\eeq 

\noindent where $\bar{p}(S) = \frac{1}{n}\sum\limits_{i}^{n} p_i$ with $p_i = expit(S_i)$. Note that $\phi^B(S)$ attains its maximum of $1$ if the sample outcome prevalence is exactly 50\%. Designs with higher values of $\phi^B(S)$ may be interpreted as having more predicted outcome class balance and therefore better learning of revision parameters.

\subsection{A computational framework to evaluate PCC design configurations}
\label{pcc_sec:methods_3implementation}

\noindent We now describe a computational framework to evaluate and select configurations $(k,w)$ for PCC designs. The proposed empirical framework is based on Monte Carlo of expected sample score information summarized with $\phi^D(S)$ and $\phi^B(S)$ under considered design configurations. Then, resulting estimated information response surfaces are visualized with pairs of contour plots, which allows for convenient evaluation of information from competing design configurations.\\

\begin{figure}[!htbp]
\centering 
\caption{Pairs of contour plots for expected sample information response surfaces calculated using D-optimality (left) and Binary Entropy (right) information functions, based on simulated scores distributed as $S \sim N(-1.5, 1)$.}
\includegraphics[width=7in,height=3in]{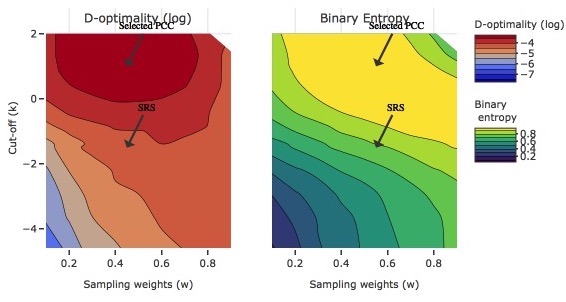} 
\label{pcc_fig:plot_pair}
\end{figure}

\noindent Details of the proposed computational framework are in Algorithm \ref{pcc_alg:comp_pcc}. First, two grids varying in cut-offs $k$ and stratum weights $w$ are specified. Then, for each considered configuration as well as under SRS, expected values of sample information functions are computed using Monte Carlo. Finally, resulting estimated information response surfaces are visualized with pairs of contour plots, where we illustrate an example in Figure \ref{pcc_fig:plot_pair}.

\setcounter{theorem}{0}
\begin{algorithm} Computational framework for PCC design configuration evaluation.
\begin{itemize}
\item[] \textbf{Input} Scores $S$; grid $k \in supp(S)$; grid $w \in [0,1]$.
\item[] \textbf{Do} Over the grids of $k \in supp(S)$, $w \in [0,1]$:
	\begin{enumerate}
    \item Draw sample of $S$ using configuration $(k,w)$.
    \item Calculate information functions $\phi^D(S|k,w)$ and $\phi^B(S|k,w)$ based on sample data.
    \item Repeat 1. and 2. a total of $B$ times.
    \end{enumerate}
\item[] \textbf{Return} Matrices of information response surface for $\phi^D(S)$ and $\phi^B(S)$.
\end{itemize} 
\label{pcc_alg:comp_pcc}
\end{algorithm}

\noindent To illustrate the information response surfaces computed from Algorithm \ref{pcc_alg:comp_pcc}, we simulated scores with distribution $S \sim N(-1.5, 1)$. Figure \ref{pcc_fig:plot_pair} shows resulting sample information response surfaces computed under the D-optimality (left) and Binary Entropy (right) information functions, where warmer colors represent better designs. For the example in Figure \ref{pcc_fig:plot_pair}, within the class of considered PCC design configurations, expected sample D-optimality $\phi^D(S)$ ranged between $-8$ and $-3$ and expected sample Binary Entropy $\phi^B(S)$ ranged between $0.20$ and $0.99$. We indicated with arrows expected sample information surfaces under SRS as well with a ``selected'' PCC design with configuration $(k,w)$ = $(1, 0.50)$, where qualitatively the ``selected'' configuration results in higher information compared to SRS as measured with D-optimality and Binary Entropy.\\

\noindent For a quantitative interpretation of information function comparisons, note that comparing $PCC^*$ having $(k,w)$ = $(1, 0.50)$ to SRS

\beqa\bal 
\dfrac{\phi^D(S|PCC^*)}{\phi^D(S|\text{SRS})} = \dfrac{-3.12}{-4.12} \:\:&\text{  and }&\:\:
\dfrac{\phi^B(S|PCC^*)}{\phi^B(S|\text{SRS})} = \dfrac{0.99}{0.77}.\\
\eal\eeqa 

\noindent Then, by exponentiating $\phi^D(S)$ the ratio of determinant of information functions comparing designs is

\beqa\bal 
\dfrac{det(\mathbf{I}_m(S|PCC^*))}{det(\mathbf{I}_m(S|SRS))} &= \dfrac{exp(-3.12)}{exp(-4.12)} \approx 2.72, 
\eal\eeqa 

\noindent and by reversing $\phi^B(S)$ the ratio of sample outcome prevalences is

\beqa\bal 
\dfrac{\bar{p}(S|PCC^*)}{\bar{p}(S|SRS)} &= \dfrac{0.49}{0.23} \approx 2.13.
\eal\eeqa 

\noindent Therefore, comparing $PCC^*$ to SRS, the expected recalibration parameter confidence region is about $2.72$ times smaller, indicating an effective sample size savings about three times. In addition, the expected sample outcome prevalence comparing $PCC^*$ to SRS is about $2.13$ higher towards outcome class balance. Selecting configurations resulting in increased information as measured by $\phi^D(S)$ and $\phi^B(S)$ ultimately improves the dual modification learning goals of recalibration power and revision support recovery, as we demonstrate through simulation in Section \ref{sec:pcc_sims}. Next, we remark on the design impact of PCC designs on modification learning.

\subsection{Design impact on modification learning}
\label{pcc_sec:methods_4properties}

\noindent To characterize design impact on modification learning, we focus on the criteria of design validity and design resource efficiency. For a given sampling design, ``validity'' for modeling refers to whether using resulting samples may provide valid inference, while ``resource efficiency'' refers to whether using resulting samples can achieve modeling goals at a lower cost compared to a ``baseline" design (usually SRS). For the modification learning problem, we describe the design validity of PCC designs and the design resource efficiency compared to SRS.

\subsubsection*{Design validity}

\noindent Design validity may be framed as a missing data problem \cite{zadrozny2004learning}, where distributional differences comparing resulting samples to the ``complete'' cohort determines whether the design is valid. Since PCC designs is a member of the general predictive score sampling class (Definition \ref{pcc_def:design_class}), sampling is based only on original model scores $S$. Denote $\Delta$ as the indicator of a subject being included in a sample. Then, for all designs within the predictive score sampling class

\beq\bal 
\Delta \perp Y|(S,\mathbf{X}),
\label{pcc_eq:design_validity}
\eal\eeq  

\noindent implying $f(Y|S,\mathbf{X},\Delta) = f(Y|S,\mathbf{X})$, therefore establishing design validity. Due to \eqref{pcc_eq:design_validity}, fitting the modification learning model \eqref{pcc_eq:model} using samples drawn with the PCC design results in asymptotically equivalent inference as if the entire cohort were available.\\

\noindent The condition \eqref{pcc_eq:design_validity} is also known as the (outcome) Missing At Random (MAR) property from the missing data literature \cite{little2014statistical}. Samples that have outcome MAR are intentionally biased from the cohort; mathematically this means that $f(\Delta|S) \neq f(\Delta)$, but any differences may be characterized. For the PCC design, sampling depends on a specific function of the score, which is whether scores exceed threshold $k$ with stratum frequency $w$. Therefore, the sampling distribution under PCC is essentially a re-weighting of that under SRS, where weights are specified within strata defined by the fixed threshold $k$ (Lemma \ref{pcc_lemma:1}). 

\setcounter{theorem}{0}
\begin{lemma} Equivalence of SRS and strata frequency weighted PCC sampling distributions.\\
For the PCC design based on stratum frequency $w$ for strata defined by scores $S$ exceeding cut-off $k$, then for fixed $k \in supp(S)$ the sampling weights for subject $i$ are

\beq\bal 
P_{PCC}(\Delta_i=1) &= \begin{cases}
P_{SRS}(\Delta_i = 1) \times \dfrac{w}{P(S_i > k)} ,& S_i > k\\
P_{SRS}(\Delta_i = 1) \times \dfrac{1-w}{P(S_i \leq k)} ,& S_i \leq k
\end{cases}, 
\label{pcc_eq:sampling_weights}
\eal\eeq 

\noindent and when $w=P(S_i>k)$,

\beqa\bal 
P_{PCC}(\Delta_i=1) &= P_{SRS}(\Delta_i=1).
\eal\eeqa 
\label{pcc_lemma:1}
\end{lemma}

\subsubsection*{Design resource efficiency}

\noindent From Lemma \ref{pcc_lemma:1}, it is clear that PCC design over-represents subjects with higher scores in resulting samples. A natural question then, is if such re-weighting provides higher sample ``information". Intuitively, stratified sampling that up-weights the more informative strata results in overall higher information. In particular, for the Binary Entropy information function $\phi^B(S)$, we show a direct correspondence between stratum weights and sample information in Lemma \ref{pcc_lemma:2}. By over-representing subjects with scores $S>k$ more than under the SRS distribution, resulting samples have higher outcome class balance, assuming that outcome probabilities $p_i$ are monotone increasing with scores $S_i$.\\

\begin{lemma} PCC design configurations for higher sample Binary Entropy.\\
For the PCC design with design configurations $(k,w)$, assume that outcome probabilities $p_i$ are monotone in scores $S$. Then, for fixed cut-off $k \in supp(S)$, using stratum weights $w$ such that $w > P(S>k)$ results in

\beqa\bal 
\phi^B(S|PCC) &> \phi^B(S|SRS).
\eal\eeqa 

\label{pcc_lemma:2}
\end{lemma}

\noindent It is also possible to show that the D-optimality function may be increased by re-weighting sample scores. However, required PCC design configurations are more complicated than simple fixed cut-offs and higher stratum weights as shown for the Binary Entropy function. As the information functions $\phi^D(S)$ and $\phi^B(S)$ are intermediaries of true modification learning goals, we omit extensive discussion, and instead remark that ordering of the D-optimality information function may be shown through the monotonicity property \cite{pukelsheim1993optimal}. To demonstrate that design configuration evaluation and selection using the PCC framework may improve on the ultimate modification learning goals of recalibration parameter testing power and revision parameter support recovery, we next provide empirical evidence through Monte Carlo simulation.

\section{Simulations}
\label{sec:pcc_sims}
\subsection{Data generating mechanism to simulate model modification learning}
\label{pcc_sec:sims_1overview}

\noindent We generated synthetic data ($n=100,000$) to demonstrate the benefit of using PCC designs for model modification learning to a new setting. True outcomes $Y$ were generated using model \eqref{pcc_eq:model} across a range of scenarios, through specifying various modification learning parameters $\alpha_0$, $\alpha_1$ and $\tilde{\gamma} \in \mathcal{R}^{100}$. Features $\tilde{X}$ and scores $S$ were considered ``fixed'' across the various data scenarios, generated using the Linear Discriminant Analysis (LDA) assumption. Conditioned on the original outcome prevalence $\pi^0$, features were generated as

\beqa\bal 
\tilde{X}^T_i |(Y^{initial}_i = y) &\sim N(\tilde{\mu}_{y}, \mathbf{\Sigma}_y),
\eal\eeqa 

\noindent with $\tilde{\mu}_{y}$ and $\mathbf{\Sigma}_y$ set such that

\beqa\bal 
logit(E[Y^{initial}_i|\tilde{X}^T_i ]) &= \beta_0 + \tilde{X}^T_i  \tilde{\beta}\\
\tilde{\beta}_j &= \begin{cases}
0.7 & j = 1, \hdots, 10 \\
-0.7 & j = 11, \hdots, 20 \\
0 & j = 21, \hdots, 100. \\
\end{cases}
\eal\eeqa

\noindent The original scores were defined as $S _i= \beta_0 + \tilde{X}^T_i \tilde{\beta}$, and the true  mean model was thus simulated as

\beqa\bal 
logit(E[Y_i|S_i, \tilde{X}^T_i]) &= \alpha_0 + \alpha_1 S_i + \tilde{X}^T_i \tilde{\gamma}
\eal\eeqa 

\noindent From each simulated cohort, samples drawn using PCC and SRS were compared based on metrics for recalibration (statistical power) and revision (support recovery). We used the same design, $\text{PCC}^{sim}$ with $(k,w) = (-1,0.50)$ for consistency across the various simulated scenarios.
\subsection{Design benefit for recalibration}
\label{pcc_sec:sims_2recalibration}

\noindent For all recalibration simulations, the power of recalibration tests was compared between using samples drawn with SRS or $\text{PCC}^{sim}$. The initial outcome prevalence was set as $\pi^0 = 0.10$, the true recalibration intercept $\alpha_0$ was set to be one of $-\log(2)$, $-\log(1.5)$, or $0$, the true recalibration slope $\alpha_1$ was set to be one of 0.6, 0.8, or 1, and the true revision parameters were set to be $\tilde{\gamma} = \tilde{0}$ (no revision). Therefore, the synthetic data simulated scenarios where the outcome is relatively rare, that outcome prevalence is expected to be even lower in the new cohort, and that the model was over-fitted in the source cohort. For each scenario and over a grid of sample sizes, the empirical power was calculated across $B=500$ simulations. These recalibration parameters were selected based on previous work illustrating the sample size requirements for model external validation studies \cite{vergouwe2005substantial}.\\

\begin{figure}[!h]
\centering 
\caption{Simulation results for model recalibration.}
\label{pcc_fig:sims_recalibration}
\subcaption[\textit{toc}]{Average empirical power of the Likelihood Ratio Test for logistic recalibration, comparing $\text{PCC}^{sim}$ with $(k,w) = (-1,0.50)$ to SRS for a range of sample sizes under various recalibration parameters, over $B=500$ simulations.}
\label{pcc_fig:sims_power}
\includegraphics[width=6.5in,height=5in]{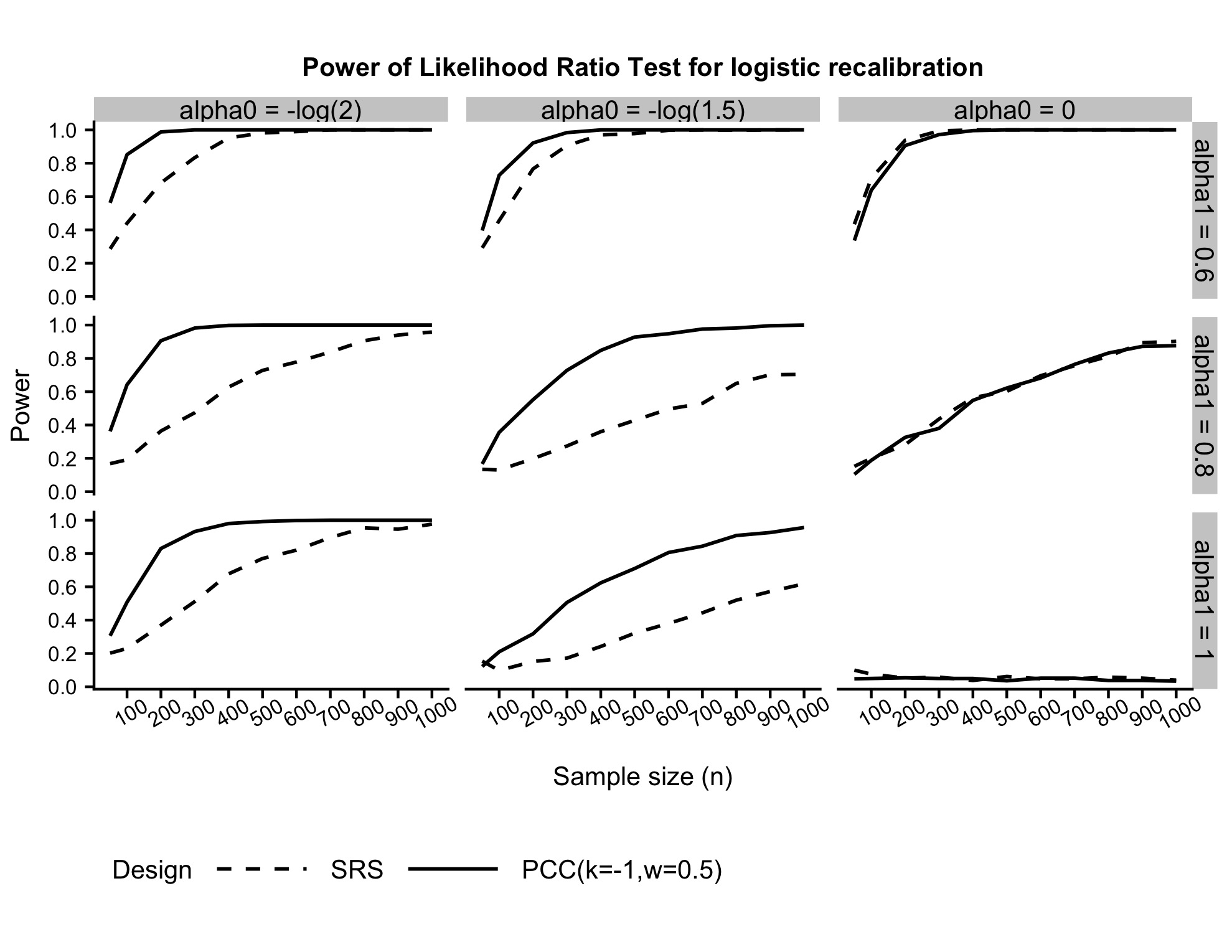}\\
\subcaption{Contour plots of D-optimality (log transformed) information functions of sample scores computed using true modification learning parameters for $(\alpha_0,\alpha_1) = (-\log(2),0.80)$ (left), $(\alpha_0,\alpha_1) = (-\log(1.5),0.60)$ (middle), and $(\alpha_0,\alpha_1) = (-\log(1.5),1)$ (right). Simulation averages were computed based on a sample size of $n=100$.}
\label{pcc_fig:sims_contour_d_optimality}
\includegraphics[width=2in,height=1.75in]{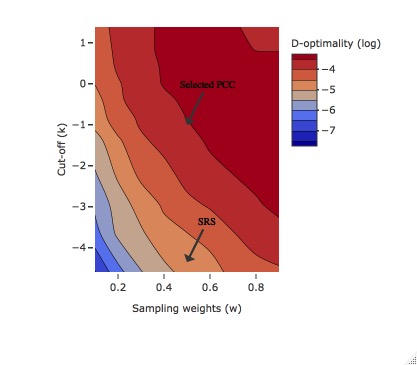}
\includegraphics[width=2in,height=1.75in]{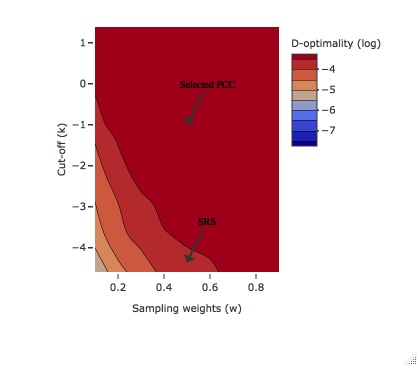}
\includegraphics[width=2in,height=1.75in]{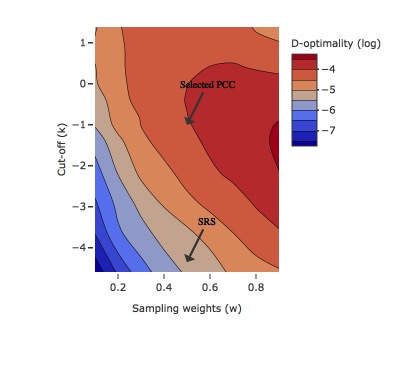}
\end{figure}

\noindent Figure \ref{pcc_fig:sims_recalibration} illustrates the simulation results for the logistic recalibration test; additional simulation results for testing the recalibration intercept and slope are shown in Appendix \ref{pcc_sec:appendix_3}. Overall, Figure \ref{pcc_fig:sims_power} shows that using $\text{PCC}^{sim}$ provided equivalent or higher power compared to SRS across the range of investigated recalibration parameters. For scenario-specific comparisons, we compare three power curves in Figure \ref{pcc_fig:sims_power} with the true D-optimality information response surfaces in Figure \ref{pcc_fig:sims_contour_d_optimality}.\\

\noindent First, consider the true data generating mechanism of $\alpha_0 = -\log(2)$ and $\alpha_1 = 0.80$ (left column middle row), where using $\text{PCC}^{sim}$ resulted in a higher power curve compared to using SRS. This observation may be explained by the D-optimality contour surfaces, where the joint confidence region for estimating the recalibration parameters was about $1.5$ times smaller using $\text{PCC}^{sim}$ compared to SRS. For true $\alpha_0 = -\log(1.5)$ and $\alpha_1 = 0.60$ (middle column top row), using $\text{PCC}^{sim}$ provided some improvement over SRS. However, the SRS power curve was already sufficiently high, as shown by the overall high D-optimality values across all PCC designs (including the SRS-equivalent configuration). For $\alpha_0 = -\log(1.5)$ and $\alpha_1 = 1$ (middle column bottom row), $\text{PCC}^{sim}$ provided some improvement over SRS but both power functions were low, again demonstrated by the overall low D-optimality values for this set of recalibration parameters. Note that when no model recalibration is needed ($\alpha_0 = 0$ and $\alpha_1 = 1$), using both SRS and $\text{PCC}^{sim}$ samples provided the correct size.\\

\noindent The power functions in Figure \ref{pcc_fig:sims_power} may be used for sample size calculation and study planning. For example, consider the scenario where true $\alpha_0 = -\log(1.5)$ and $\alpha_1 = 0.80$ (middle column middle row), to achieve 80\% power for the logistic recalibration test requires close to $n=1000$ abstracted samples when using SRS. For an outcome with 10\% prevalence this corresponds to an effective sample size of abut $100$, comparable to what was found in \cite{vergouwe2005substantial}. However, to achieve this same power using $\text{PCC}^{sim}$ only requires the abstraction of $n=200$ samples, a cost savings of almost five-fold. 

\clearpage 

\subsection{Design benefit for revision}
\label{pcc_sec:sims_3revision}

\noindent For all revision simulations, the average False Discovery Rate (FDR) and False Exclusion Rate (FER) \eqref{pcc_eq:fdr_fer} in selecting truly predictive features were compared between using SRS and PCC. The original outcome prevalence was set as either $\pi^0 = 0.10$ or $\pi^0 = 0.25$, the true recalibration parameters as $\alpha_0 = -\log(3)$ and $\alpha_1 = 0.90$, and the true revision parameters with effect size $|\gamma_j| = 0.60$ and sparsity ratio (proportion of non-zero revision parameters) of $f = 0.05$. The simulated data compares the effect of using PCC designs on support recovery measures for data with two different outcome prevalences. For each scenario and over a grid of sample sizes, estimated revision parameters $\hat{\gamma}(\lambda)$ were fitted using \eqref{pcc_eq:lasso_procedure} for the solution path defined by $\lambda \in [-\log(8),-\log(2)]$. Resulting estimates were compared against true revision parameters, and empirical FDR and FER calculated across $B=500$ simulations.\\

\begin{figure}[!htbp]
\centering 
\caption{Simulation results for model revision.}
\label{pcc_fig:sims_revision}
\subcaption{Empirical average False Exclusion Rate (FER) versus False Discovery Rate (FDR) comparing $\text{PCC}^{sim}$ with $(k,w) = (-1,0.50)$ to SRS for a range of sample sizes, over $B=500$ simulations.}
\label{pcc_fig:sims_support_recovery}
\includegraphics[width=6.5in,height=4in]{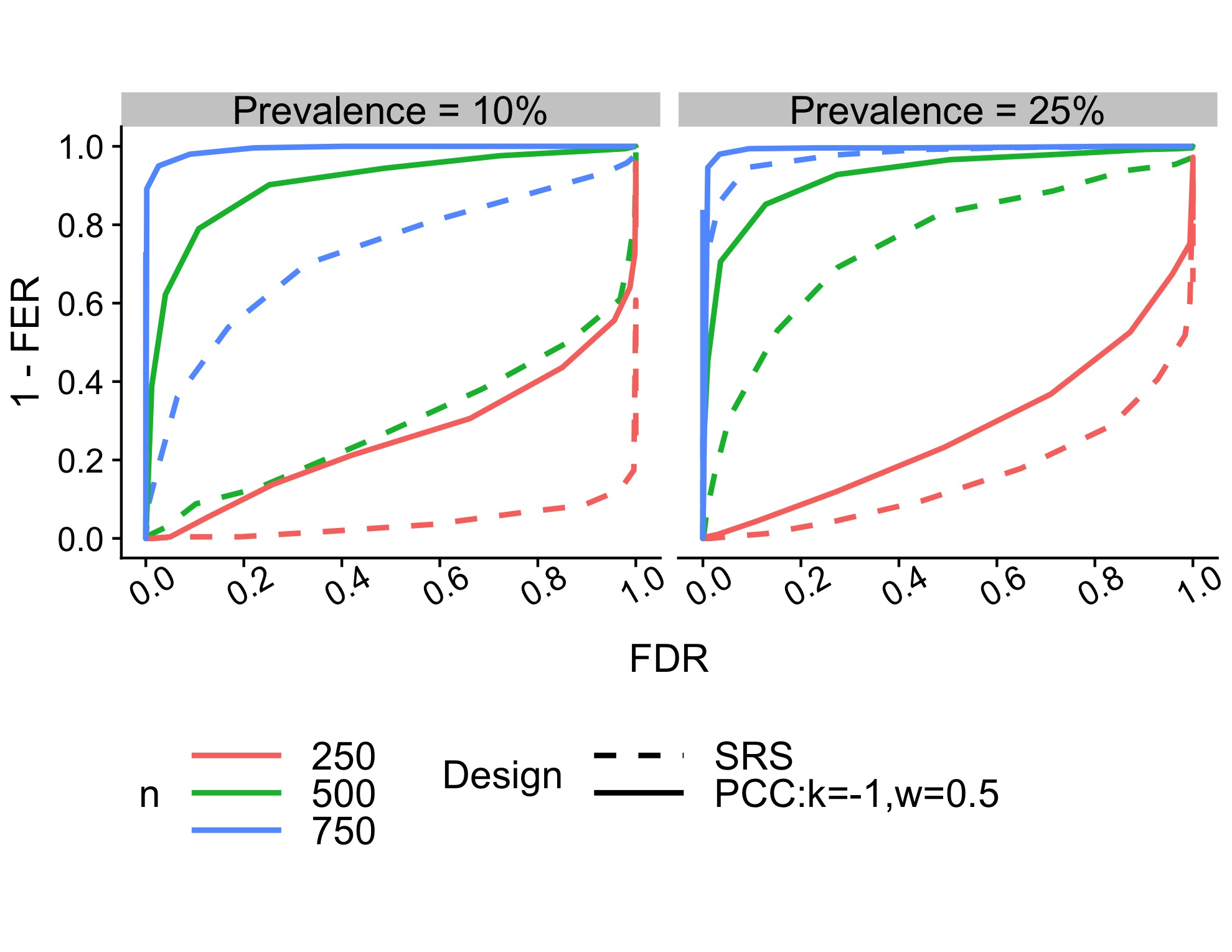}
\subcaption{Contour plots of Binary Entropy information functions of sample scores computed using true modification learning parameters for outcome prevalence of $\pi^0 = 0.10$ (left) and $\pi^0 = 0.25$ (right). Simulation averages were computed based on a sample size of $n=300$.}
\includegraphics[width=2in,height=2in]{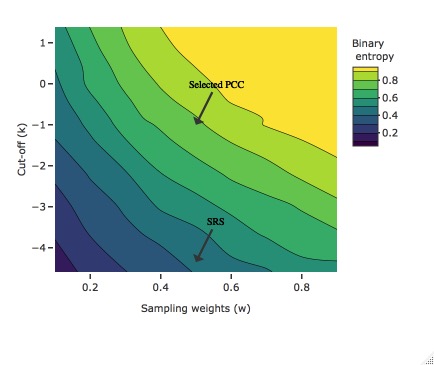}
\includegraphics[width=2in,height=2in]{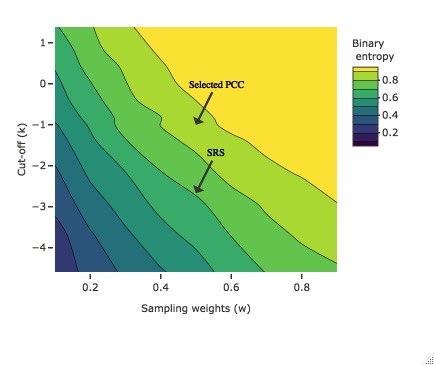}
\label{pcc_fig:sims_contour_binary_entropy}
\end{figure}

\noindent Figure \ref{pcc_fig:sims_revision} illustrates the simulation results. Figure \ref{pcc_fig:sims_support_recovery} illustrates the simulated average 1-False Exclusion Rate (1-FER) versus False Discovery Rate (FDR), where curves towards the top-left corner of the plot indicate better designs. For each sample size (n = 250, 500, 750) and each sampling design (SRS or PCC), the point at $(0,0)$ indicates support recovery of model coefficients estimated at a fixed high penalty, where both the false positives and true positives are zero since all coefficients are excluded. As penalty decreases, coefficients enter the solution path, increasing both false positives and true positives in estimating true model coefficients. Finally, when all coefficients are included in the model both false positives and true positives tend to 1.\\

\noindent For each sample size, we may compare resulting support recovery curves between $\text{PCC}^{sim}$ and SRS. For an outcome prevalence of 10\%, when $n=250$, using SRS results in models that exclude almost all predictive features, unless all features were selected, but $\text{PCC}^{sim}$ may allow recovery of about 20\% truly predictive features at FDR=0.40. Support recovery improved with sample size, where the improvement was faster for $\text{PCC}^{sim}$: at $n=500$ for FDR=0.20 using $\text{PCC}^{sim}$ recovers 80\% while SRS only recovers 10\% of truly predictive features, and at $n=750$ using $\text{PCC}^{sim}$ results in almost perfect support recovery, but using SRS may still result in false negatives and false positives. Similar patterns are observed for the 25\% outcome prevalence scenario, where for every sample size support recovery curves using $\text{PCC}^{sim}$ were higher compared to SRS, although separation of $\text{PCC}^{sim}$/SRS support recovery curves were smaller compared to the 10\% prevalence scenario.\\

\noindent To explain this observation, we turn to the sample Binary Entropy as illustrated in Figure \ref{pcc_fig:sims_contour_binary_entropy}. For the 10\% outcome prevalence, sample Binary Entropy values for SRS and $\text{PCC}^{sim}$ were $0.43$ (sample outcome prevalence = $9\%$) and $0.80$ (sample outcome prevalence = $25\%$) respectively. For the 25\% outcome prevalence, sample Binary Entropy values for SRS and $\text{PCC}^{sim}$ were $0.77$ (sample outcome prevalence = $23\%$) and $0.85$ (sample outcome prevalence = $28\%$) respectively. Thus, while using the specified $\text{PCC}^{sim}$ design more than doubled sample outcome prevalence for the 10\% prevalence scenario, it only slightly increased outcome class balance for the 25\% prevalence scenario. Such differences in sample Binary Entropy values corroborated the support recovery curves illustrated in Figure \ref{pcc_fig:sims_support_recovery}. Therefore, through selecting samples targeted towards higher Binary Entropy, $\text{PCC}^{sim}$ designs improved support recovery of predictive features - such effect was more pronounced for rarer outcomes.

\clearpage 

\subsection{Design local robustness}
\label{pcc_sec:sims_4robust}

\noindent Our simulations thus far demonstrated the benefit of using $\text{PCC}^{sim}$ for model recalibration and revision, where design configurations were evaluated conditioned on true modification learning parameters. Here, we investigate by simulation whether the local robustness to parameter mis-specification but assuming model \eqref{pcc_eq:model}.\\

\begin{figure}[!htbp]
\centering 
\caption{Contour plots of sample scores summarized with the  D-optimality (log) information function, based on data generated based on model \eqref{pcc_eq:model} using Linear Discriminant Analysis (LDA) features with $\pi^0 = 0.10$.}
\label{pcc_fig:contour_recalibration}

\subcaption{D-optimality contour surfaces computed using the  assumption $\alpha_0=0$, $\alpha_1=1$, $\tilde{\gamma} = \tilde{0}$.}
\label{pcc_fig:contour_recalibration_guess}
\includegraphics[width=3.5in,height=2.5in]{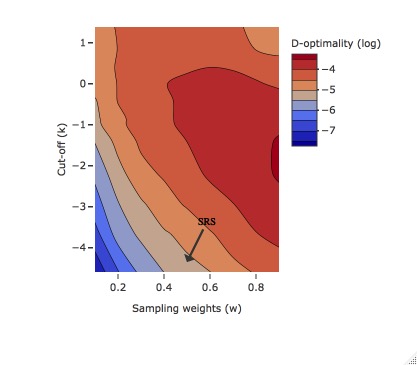} 

\subcaption{D-optimality contour surfaces computed using true $(\alpha_0,\alpha_1) = (-\log(2),1)$ (left); $(\alpha_0, \alpha_1) = (0, 0.8)$ (middle); $(\alpha_0, \alpha_1) = (0, 1.25)$ (right). The true revision parameters were $\tilde{\gamma} = \tilde{0}$ for all scenarios.}
\label{pcc_fig:contour_recalibration_true}
\includegraphics[width=2.05in,height=1.75in]{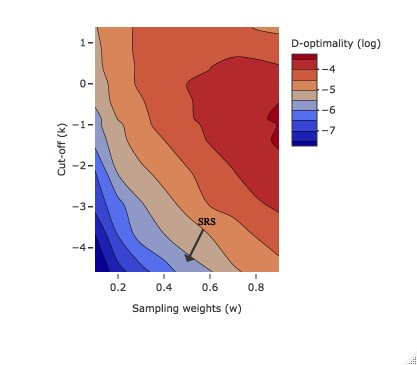} 
\includegraphics[width=2.05in,height=1.75in]{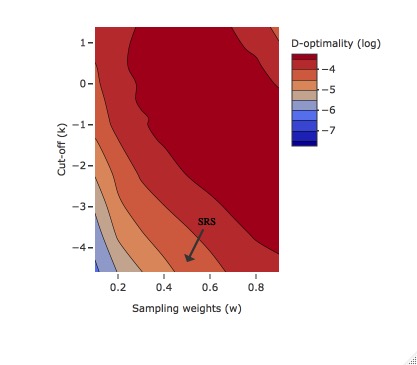}
\includegraphics[width=2.05in,height=1.75in]{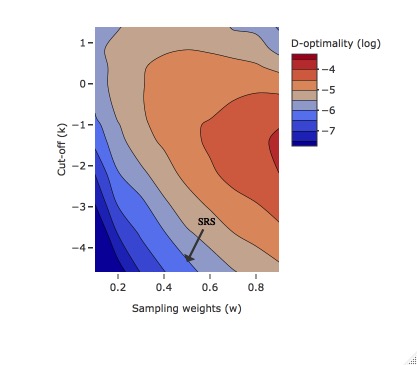}
\end{figure}

\begin{figure}[!htbp]
\centering 
\caption{Contour plots of sample scores summarized with the  Binary Entropy information function, based on data generated based on model \eqref{pcc_eq:model} using Linear Discriminant Analysis (LDA) features with $\pi^0 = 0.10$.}
\label{pcc_fig:contour_revision}

\subcaption{Binary Entropy contour surfaces computed using the  assumption $\alpha_0=0$, $\alpha_1=1$, $\tilde{\gamma} = \tilde{0}$.}
\label{pcc_fig:contour_revision_guess}
\includegraphics[width=3.5in,height=2.5in]{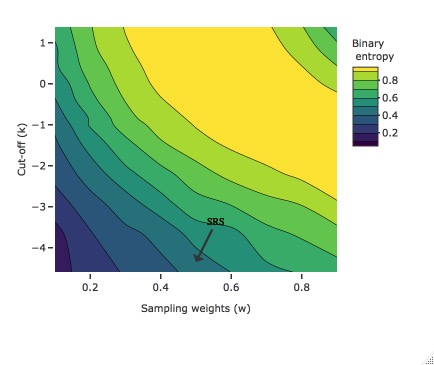} 

\subcaption{Binary Entropy contour surfaces computed using using true $(\alpha_0,\alpha_1) =(0,1)$, $|\gamma_j| = 0.60$ (left); $(\alpha_0,\alpha_1) =(-\log(3), 0.90)$, $\tilde{\gamma}$, $|\gamma_j| = 0.60$ (middle); $(\alpha_0,\alpha_1) =(0,1)$, $|\gamma_j| = 1.50$ (right). f=5\% of the revision parameters are non-zero with effect size $|\gamma_j|$; remaining revision parameters are non-predictive.}
\label{pcc_fig:contour_revision_true}
\includegraphics[width=2.05in,height=2in]{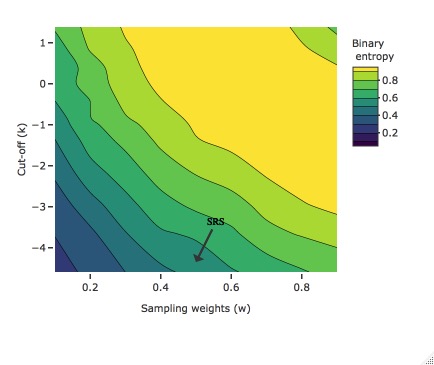} 
\includegraphics[width=2.05in,height=2in]{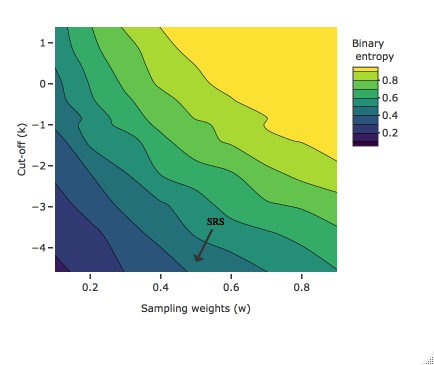} 
\includegraphics[width=2.0in,height=2in]{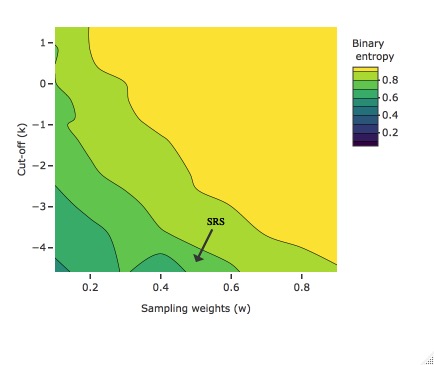} 
\end{figure}

\noindent Figure \ref{pcc_fig:contour_recalibration} shows sample D-optimality contour surfaces, where configurations with warmest colors (red) provide highest sample D-optimality values. Note that warmest color regions coincide regardless of whether computations were based on the null assumption of $\alpha_0=0$, $\alpha_1=1$, $\tilde{\gamma} = \tilde{0}$ (Figure \ref{pcc_fig:contour_recalibration_guess}) or under the actual recalibration parameters (Figure \ref{pcc_fig:contour_recalibration_true}). When true $(\alpha_0, \alpha_1)$ deviates from $(0,1)$, systematic shifting of D-optimality values were observed, where contour surfaces were shifted upwards when true $\alpha_0 < 0$, rotated counter-clockwise when true $\alpha_1 \in [0,1]$, and rotated clockwise when true $\alpha_1 > 1$. However, conclusions based on sample score information computed under any considered assumptions are comparable, where for the illustration in Figure \ref{pcc_fig:contour_recalibration} designs to the right of the contour plots indicate high information designs. Figure \ref{pcc_fig:contour_revision} shows a similar story for sample scores summarized with the Binary Entropy information function, where contour plots were systematic shifted, rotated, and/or scaled but overall conclusions are locally robust to slight parameter mis-specifications.\\

\begin{figure}[!htbp]
\centering 
\caption{Effect of score distribution on D-optimality and Binary Entropy information functions of sample scores. Data generating mechanisms are: scores based on LDA features with $\pi^0 = 0.10$ (left); LDA features with $\pi^0 = 0.50$ (middle); $S \sim N(1.5, 1^2)$ (right). All contour surfaces are computed based on the assumption $\alpha_0=0$, $\alpha_1=1$, $\tilde{\gamma} = \tilde{0}$. LDA: Linear Discriminant Analysis; $\pi^0$: the original outcome prevalence assumed in the source cohort.}
\label{pcc_fig:contour_score_dist}

\subcaption{Score distributions in the simulated cohort.}
\label{pcc_fig:score_dist}
\includegraphics[width=2.05in,height=2in]{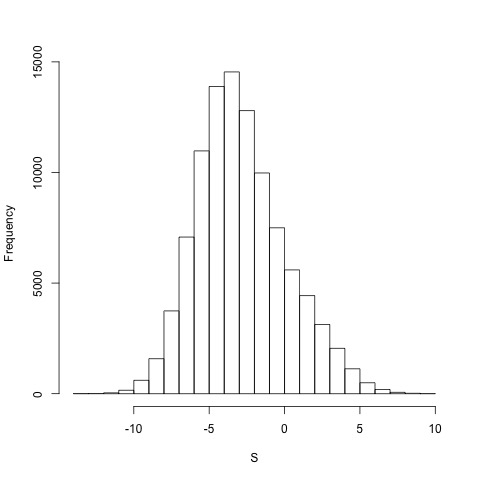}
\includegraphics[width=2.05in,height=2in]{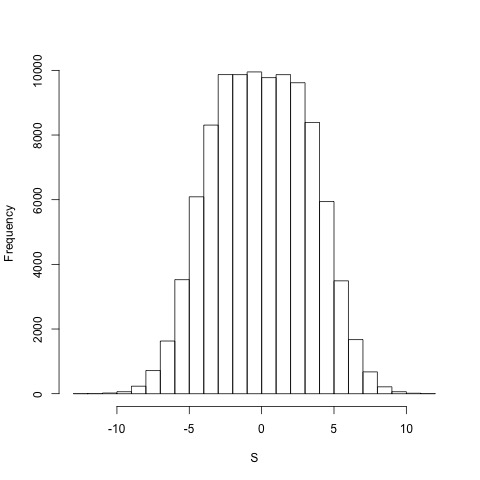}
\includegraphics[width=2.05in,height=2in]{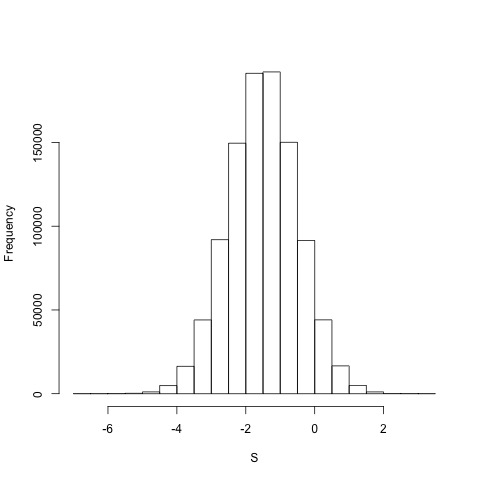}

\subcaption{D-optimality (log) contour surfaces.}
\label{pcc_fig:dopt_score_dist}
\includegraphics[width=2.05in,height=1.75in]{Figures/pcc_contour_recalib_guess.jpeg}
\includegraphics[width=2.05in,height=1.75in]{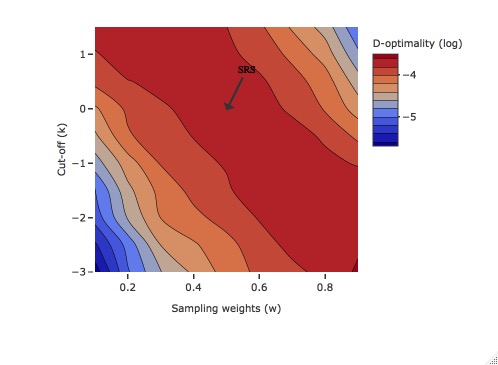} 
\includegraphics[width=2.05in,height=1.75in]{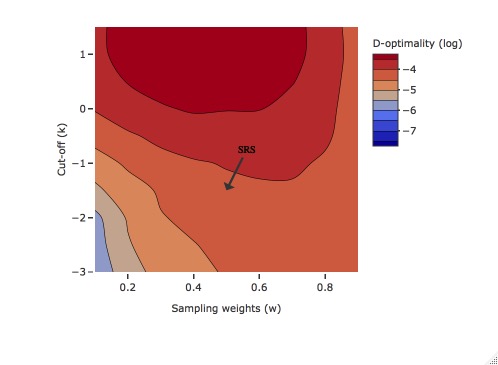} 

\subcaption{Binary Entropy contour surfaces.}
\label{pcc_fig:binent_score_dist}
\includegraphics[width=2.05in,height=2in]{Figures/pcc_contour_revision_guess.jpeg} 
\includegraphics[width=2.05in,height=2in]{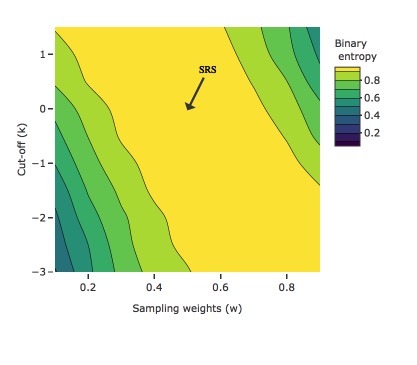} 
\includegraphics[width=2.05in,height=2in]{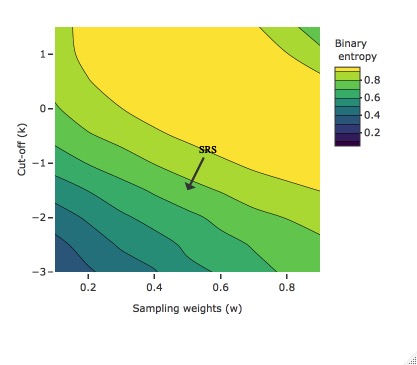} 
\end{figure}

\noindent However, in contrast to mis-specification to parameter mis-specification, score distributions may affect design evaluation conclusions. Figure \ref{pcc_fig:contour_score_dist} illustrates the effect of using different score distributions (Figure \ref{pcc_fig:score_dist}) on sample scores summarized with the D-optimality (Figure \ref{pcc_fig:dopt_score_dist}) and Binary Entropy (Figure \ref{pcc_fig:binent_score_dist}) information functions. Depending on cohort score distribution, high information designs as evaluated by D-optimality and Binary Entropy differ. However, such conclusions are less concerning, as true score distributions are assumed to be known for all subjects in the new cohort.

\clearpage

\clearpage

\clearpage

\section{Illustration: Modification learning on radiology report}
\label{sec:pcc_data}
\subsection{Radiology reports and modification learning across imaging modalities}

\noindent To illustrate the proposed PCC design on reducing sample size requirements for modification learning, we consider radiology reports arising from different imaging modalities. Radiology reports constitute the formal communication of imaging study results by trained radiologists, contain important information about radiographic findings, but often presented as unstructured data via free-text. Therefore, the collection of radiographic findings outcome labels require abstraction by highly-trained human clinical experts, which is a labor intensive and costly process. Such costs motivate using modification learning modeling and resource efficient sampling approaches for data collection.\\

\noindent The application data set comes from radiology reports derived from the Lumbar Imaging with Reporting of Epidemiology (LIRE) study \cite{jarvik2015lumbar}. The LIRE study was a randomized pragmatic clinical trial that studied the effect of radiology report content on subsequent treatment decisions. Adult subjects were considered for study inclusion if their primary care provider (PCP) ordered either an x-ray or Magnetic Resonance (MR) imaging test of the lumbar spine. A finding of interest was vertebral fracture, which is visible on both x-ray and MR imaging modalities. To modify a ``source'' model for fracture previously developed using x-ray reports for application to MR reports, a sample of ``new'' outcome labels needs to be abstracted from MR reports. Towards the resource efficient collection of such a sample, we illustrate the benefit of using PCC designs compared to SRS, as evaluated based on the dual modification learning goals of model recalibration and model revision.

\subsection{Illustration set-up}

\noindent Our approach for illustration is through simulation on the full data, defining true source and modification learning models based on all available data, and demonstrating design effects through sub-sampling. Our illustrations require text processing and classification modeling, where we used existing routines from the \texttt{quanteda} and \texttt{glmnet} packages in \texttt{R}, respectively. To reduce any potential between-site variation, we restricted analysis to the single largest site from the LIRE study. For all subjects from this single site, we obtained a large number of simulated abstracted outcomes using a rule-based natural language processing algorithm \cite{tan2018comparison}. A total of $N^0 = 158,405$ labels (prevalence = 10\%) were obtained from source x-ray reports, and $N^1 = 41418$ labels (prevalence = 7\%) were possible from MR reports. Full details of the source and target model definition procedures are in Appendix \ref{pcc_sec:appendix_4}; here we provide an overview of the feature engineering, source model definition, and target model definition processes:\\

\noindent \textbf{Feature engineering:} Features $\tilde{X}^T_i$ were created for all x-ray and MR reports and included the following for a total of $p=131$ features:
\bi 
\item \underline{Text features:} Bag-of-words stemmed unigrams indicators, excluding rare (occurring in $<5\%$ of documents) and common (occurring in $>80\%$ of documents) terms.
\item \underline{Demographic features:} Gender (male or female) and age category ($<40$, $40$-$60$, $>60$ years).
\ei 

\noindent \textbf{Source model definition:} The source classification model was estimated using the Logistic Lasso procedure with 10-fold cross-validation based on an empirical AUC loss function. Model development was based on a random sample of $n^1=5000$ drawn from $N^1$, where $23$ non-zero coefficients were selected.\\

\noindent \textbf{Modification learning model definition:} To define the modification learning model, first scores were computed by applying the source model to features from the MR cohort. Then, modification learning parameters were estimated as follows:
\bi 
\item \underline{Recalibration parameters $\alpha_0$ and $\alpha_1$:} Estimated by fitting the modification learning model \eqref{pcc_eq:model} conditioned on no revision ($\tilde{\gamma} = \tilde{0}$).
\item \underline{Revision parameters $\tilde{\gamma}$:} Estimated by fitting modification learning model \eqref{pcc_eq:model} with the Lasso procedure \eqref{pcc_eq:lasso_procedure} to penalize features but not scores.
\ei 
\noindent This results in a ``true" modification learning model with parameters
\beqa\bal 
logit(E[Y|S,\mathbf{X}]) &= -1.03 + 0.89S+ \mathbf{X}\tilde{\gamma}
\eal\eeqa

\noindent where $6$ features in $\tilde{\gamma}$ were estimated to be non-zero: stemmed unigrams sublux = 1.42; deform = 1.15; fractur = 0.78; scoliosi = 0.78; normal = -0.53; desicc = -0.57.

\subsection{Evaluating and selecting a PCC design configuration}

\noindent Using the proposed computational framework (Algorithm \ref{pcc_alg:comp_pcc} in Section \ref{pcc_sec:methods_3implementation}), we evaluate and select, score cut-off $k$ and stratum weights $w$ that increase expected sample information within PCC class design configurations. Using original scores computed for MR reports based on the source x-ray model, we calculated resulting D-optimality \eqref{pcc_eq:d_optimality} and Binary Entropy \eqref{pcc_eq:binary_entropy} information functions with Monte Carlo.\\

\begin{figure}[!htbp]
\caption{Pairs of contour plots for D-optimality and Binary Entropy information functions of sample scores.}
\centering 
\label{pcc_fig:data_analysis_contours}
\includegraphics[width=6in,height=2in]{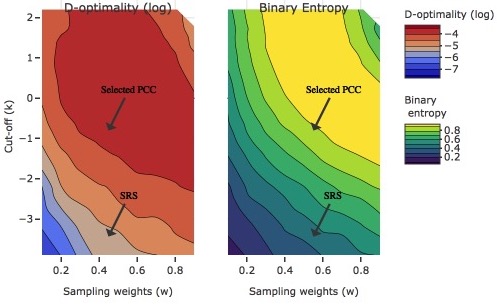}
\end{figure}

\noindent The pairs of contour plots of information functions are shown in Figure \ref{pcc_fig:data_analysis_contours}. As illustrated, there were a range of PCC design configurations that may provide improved information for model modification learning compared to using SRS. However, for this practical illustration we were additionally constrained by the total available sample size. For example, while using the configuration $(k,w)$ = $(2,0.50)$ may be ideal based purely on values of information functions, there were only $105$ subjects with scores exceeding $k=2$ resulting in a maximum sample size of $202$. Alternatively, as there are $461$ subjects with scores exceeding $k=0.80$, the configuration of $(k,w)$ = $(0.80,0.50)$ is potentially more reasonable as it allows up to a maximum of $5526$ subjects to be selected. Therefore, considering both potential design enrichment as well as practical constraints, we selected $\text{PCC}^{illus}$ with $(k,w)$ = $(0.80, 0.50)$ to illustrate the benefit of using PCC over SRS through sub-sampling.

\subsection{Benefit of PCC for modification learning}

\noindent The selected $\text{PCC}^{illus}$ was then compared to SRS through sub-sampling, with statistical power and support recovery as evaluation metrics for recalibration and revision, respectively. Over a grid of sample sizes, sub-samples of MR reports were drawn with either SRS or $\text{PCC}^{illus}$. Recalibration comparisons were based on the empirical power of likelihood ratio tests for the recalibration intercept, slope, and logistic recalibration computed over the sub-samples, fitting the modification learning model \eqref{pcc_eq:model} conditioned on no revision ($\tilde{\gamma} = \tilde{0}$). Since the true recalibration parameters were estimated to be $\alpha_0 = -1.03$ and $\alpha_1 = 0.89$ we expected empirical power to approach 1 as sample size increases, but the rate of power increase may differ depending on sampling design.\\

\noindent Revision comparisons were based on empirical support recovery curves. However, as some estimated revision parameters had small coefficients, we used alternative definitions for FDR and FER, where

\beq\bal 
FDR^{alt} &= 1-P(\boldsymbol{\hat{\gamma}_S} \in \boldsymbol{\gamma_{H}^*} \cup \boldsymbol{\gamma_{L}^*} ) \\
FER^{alt} &= 1-P(\boldsymbol{\hat{\gamma}_{S^c}} \in \boldsymbol{\gamma_{L}^*} \cup \boldsymbol{\gamma_{S^c}^*}).
\label{pcc_eq:fdr_fer_alt}
\eal\eeq 

\noindent In \eqref{pcc_eq:fdr_fer_alt}, estimated coefficients from the sub-samples $\boldsymbol{\hat{\gamma}_S}$ and $\boldsymbol{\hat{\gamma}_{S^c}}$ were defined similarly as with convention. However, true coefficients estimated from the full cohort were divided into high, low, and no signal categories $\boldsymbol{\gamma_{H}^*}$, $\boldsymbol{\gamma_{L}^*}$ and $\boldsymbol{\gamma_{S^c}^*}$, with

\beq\bal 
\boldsymbol{\gamma_H^*} &= \{\gamma_j^*: |\gamma_j^*| > 0.50 \} \\
\boldsymbol{\gamma_L^*} &= \{\gamma_j^*: 0 < |\gamma_j^*| \leq 0.50 \} \\
\boldsymbol{\gamma_{S^c}^*}&= \{\gamma_j^*: |\gamma_j^*| = 0 \}.
\label{pcc_eq:true_revision_param_alt}
\eal\eeq  

\noindent In \eqref{pcc_eq:true_revision_param_alt}, $FDR^{alt}$ is defined similar to the usual $FDR$: any feature that was selected but had a truly non-zero coefficient was considered a false positive. However using $FER^{alt}$ allows features with ``low signal''($\boldsymbol{\gamma_L^*}$) to be excluded yet not count as a false negative, therefore avoiding penalizing the non-selection of low signal features which could result in artificially high FER. The threshold $0.50$ in \eqref{pcc_eq:true_revision_param_alt} was chosen to keep the feature sparsity ratio around $f=5\%$.\\

\begin{figure}[!htbp]
\caption{Results from data example illustration.}
\label{pcc_fig:data_analysis}
\centering 
\subcaption{Empirical power of Likelihood Ratio Tests comparing $\text{PCC}^{illus}$ to SRS over $B=500$ simulations, where true recalibration and slope were respectively $\alpha_0|(\tilde{\gamma} = \tilde{0}) = -1.03$ and $\alpha_1|(\tilde{\gamma} = \tilde{0}) = 0.89$.}
\label{pcc_fig:data_analysis_recalibration}
\includegraphics[width=6in,height=3in]{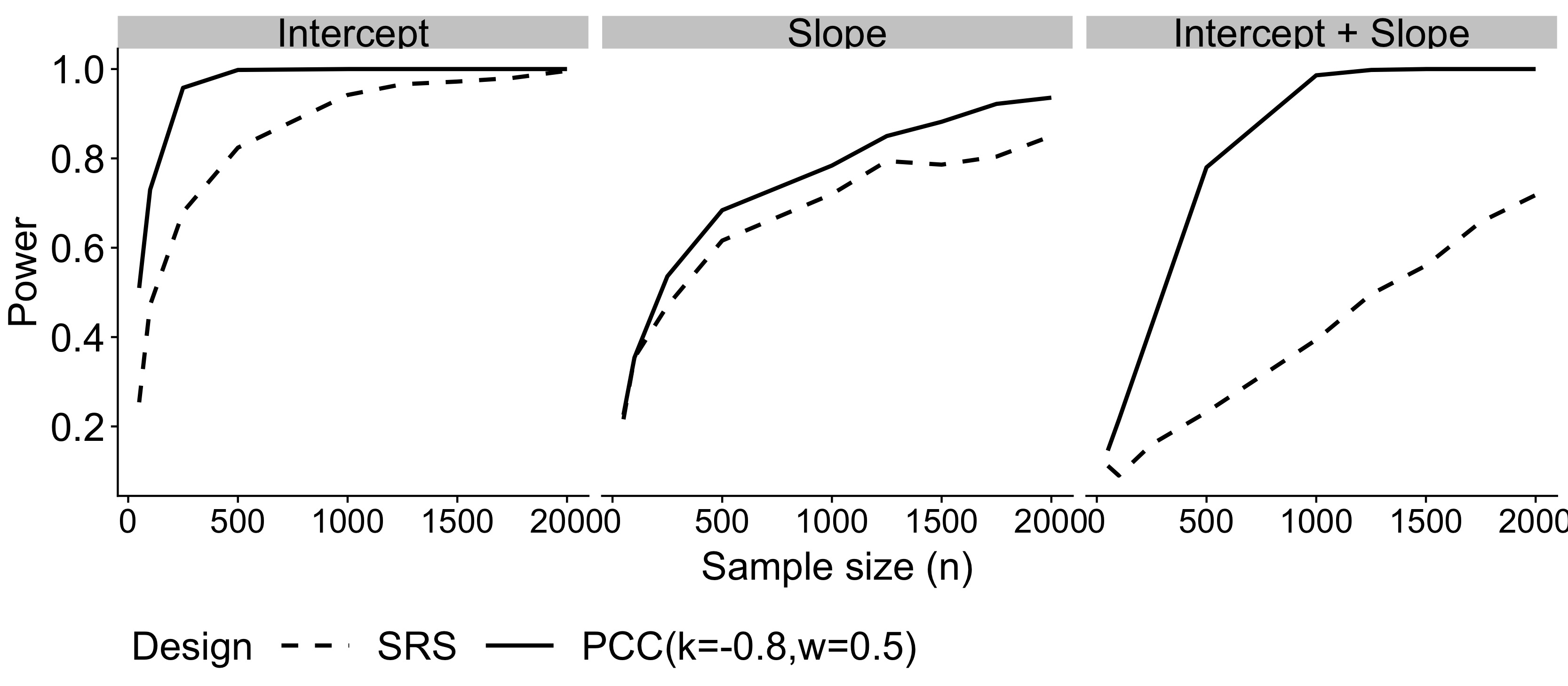} 

\subcaption{Empirical support recovery curves comparing $\text{PCC}^{illus}$ to SRS over $B=500$ simulations, using the alternative definitions of False Discovery Rate ($FDR^{alt}$) and False Exclusion Rate ($FER^{alt}$). The revision parameter vector length was $p=131$, of which $6$ features were truly non-zero.}
\label{pcc_fig:data_analysis_revision}
\includegraphics[width=4in,height=3in]{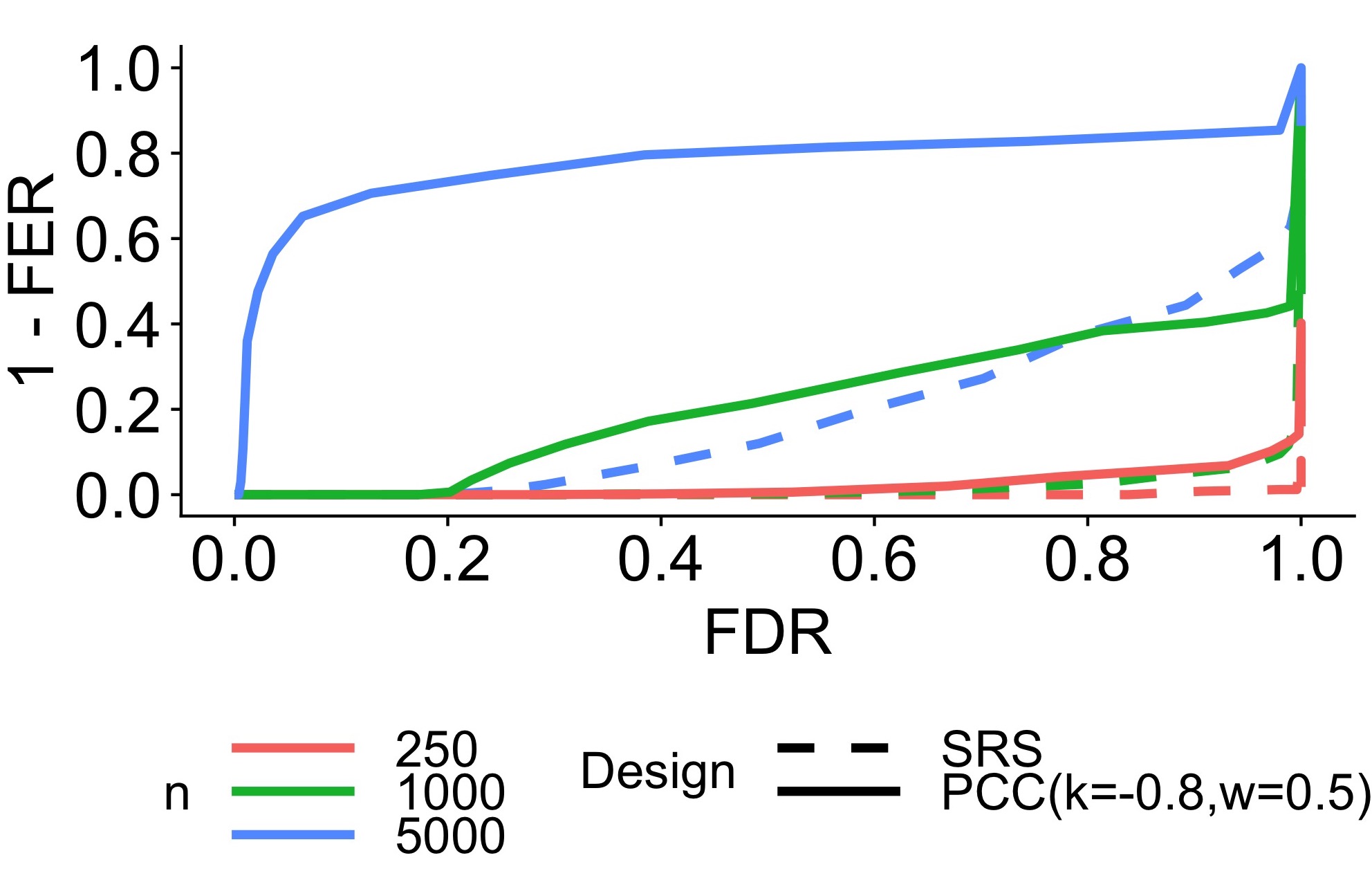} 
\end{figure}

\noindent Figure \ref{pcc_fig:data_analysis_recalibration} shows the design effect on recalibration statistical power. As illustrated, the power functions for all three recalibration tests were higher under $\text{PCC}^{illus}$ compared to under SRS. For example, to detect mis-calibration of average predictions (recalibration intercept test) with a $80\%$ power controlling test size at $0.05$, using SRS requires $n=500$ but using $\text{PCC}^{illus}$ only requires $n=250$. Therefore, for this specific external validation study, using $\text{PCC}^{illus}$ was about twice as cost efficient compared to using SRS.\\

\noindent Figure \ref{pcc_fig:data_analysis_revision} shows the design effect on revision support recovery. For the three illustrated sample sizes, using $\text{PCC}^{illus}$ resulted in better support recovery compared to using SRS. For a sample size of $n=250$, support recovery was low regardless of sampling design. For $n=1000$, while using SRS did not provide any meaningful support recovery, using $\text{PCC}^{illus}$ may recover 30\% of truly predictive features for FDR=0.50. For $n=5000$, using $\text{PCC}^{illus}$ results in the recovery of 70\% truly predictive features for FDR=0.10, however when using SRS both FDR and FER remained very high. These illustrations demonstrate that, even when the abstraction budget can be as high as $n=5000$, data collection using SRS could still lead to inaccurate model revision as measured by support recovery. 

\clearpage 

\section{Discussion}
\label{sec:pcc_discussion}

\noindent We demonstrated that using design-based principles for new outcome label collection can substantially reduce the sample size requirement for model modification learning. The proposed class of designs is based on sampling using original model predicted scores, which we showed to be amenable for valid analysis. By stratified sampling on strata defined by score values and over-representing subjects with ``high information'' scores, the resulting Predictive Case Control (PCC) design is additionally resource efficient for model recalibration and model revision. We developed a computational framework to visualize and compare design configurations within the PCC class.\\

\noindent The proposed PCC sampling design is easy to communicate and implement in practice. Intuitively, over-representing subjects with higher scores can be interpreted as over-including ``likely cases'', establishing an equivalence of the proposed method and the well-known case-control study design. Implementing the stratified PCC design is straightforward, as it only depends on two design configurations: score cut-off $k$ and stratum weights $w$. Design configuration selection is based on information functions of the scores which are directly related to the dual modification learning goals. Our proposed framework worked well for scenarios even with relatively small effect sizes for recalibration tests ($\alpha_0 \approx -\log(1.5)$, $\alpha_1 \approx 0.80$) and relatively rare outcomes (5\% - 10\%).\\

\noindent The simple alternative to PCC is to use SRS, which while results in a representative sample for external validation, may not provide sufficient effective sample sizes. More sophisticated alternatives include active and transfer learning procedures. However, since such machine-learning procedures are specifically developed towards resource efficient learning, resulting labeled samples obtained through such procedures may not be amenable for valid analysis. We remark that it is possible to view the described modification learning model as a special formulation of transfer learning, and the proposed PCC design as a pre-specified sampling design for that transfer learning task.\\

\noindent There are a few limitations of the proposed PCC design. First, we assumed that initial scores are easily computed in the new setting though applying the developed model to the ``new'' features. However, in practice it is possible that the features used for the original model development are missing in the new setting due to reasons such as incomplete data capture in the EMR. The extent of the impact of missing features on our proposed method is yet unexplored. Our simulation results showed that targeting samples for D-optimality and Binary Entropy improves the true model updating goals of recalibration and revision. While the relationship between D-optimality and statistical power is known in the statistical literature \cite{wald1943tests}, theoretical relationships between Binary Entropy and support recovery are relatively unexplored.\\

\noindent Future work directions include the translation of PCC for use in study design planning, for example sample size calculation through user-supplied parameter values. Towards that end, we plan on integrating simulation calculations and graphical displays for information response surfaces, power functions and support recovery curves into an interactive \texttt{RShiny} application: this process is under development at time of this writing. In addition, it may be interesting to explore using alternative information functions for PCC design configuration selection on the resource efficiency of modification learning. For example, the D-optimality information function may be replaced by other ``true'' information functions such as A-optimality or G-optimality that are more focused on average and prediction variance reduction, and Binary Entropy information function by other measures of outcome class balance. Ultimately, we hope to inspire design-based thinking for ``new'' outcome label collection, especially when using SRS dictates insurmountable sample size requirements.

\section{Appendix}
\label{sec:pcc_appendix}
\subsection{Proof of Lemma \ref{pcc_lemma:1}}
\label{pcc_sec:appendix_1}

\noindent For subject $i$ let

\beqa\bal 
\Delta_i &= \begin{cases} 1 ,& \text{sampled}  \\
0,& \text{otherwise}
\end{cases}
\eal\eeqa 

\noindent Under the PCC design, for strata defined by scores $S$ exceeding threshold $k$, sampling is based on re-weighting strata frequencies. Thus by construction

\beqa\bal 
P(S_i > k|\Delta_i = 1) &= w \\
P(S_i \leq k|\Delta_i = 1) &= 1-w.
\eal\eeqa 

\noindent Therefore,

\beqa\bal 
P_{PCC}(\Delta_i=1) &= \begin{cases}
P(\Delta_i = 1|S_i > k) ,& S_i > k\\
P(\Delta_i = 1|S_i \leq k) ,& S_i \leq k
\end{cases} \\
&= \begin{cases}
\dfrac{P(S_i > k|\Delta_i = 1)P(\Delta_i = 1)}{P(S_i > k)} ,& S_i > k\\
\dfrac{P(S_i \leq k|\Delta_i = 1)P(\Delta_i = 1)}{P(S_i \leq k)} ,& S_i \leq k
\end{cases} \\
&= \begin{cases}
P(\Delta_i = 1) \times \dfrac{w}{P(S_i > k)} ,& S_i > k\\
P(\Delta_i = 1) \times \dfrac{1-w}{P(S_i \leq k)} ,& S_i \leq k
\end{cases}  
\eal\eeqa 

\noindent Let $w = P(S_i>k)$. Then,

\beqa\bal 
P_{PCC}(\Delta_i=1) &= \begin{cases}
P(\Delta_i = 1) \times \dfrac{w}{w} ,& S_i > k\\
P(\Delta_i = 1) \times \dfrac{1-w}{1-w} ,& S_i \leq k
\end{cases} \\
&:= P_{SRS}(\Delta_i=1)
\eal\eeqa 

\newpage 
\subsection{Proof of Lemma \ref{pcc_lemma:2}}
\label{pcc_sec:appendix_2}

\noindent Define $p_i = \text{expit}(S_i)$ so $p$ is monotone increasing in $S$. Since PCC is a stratified sampling strategy,

\beqa\bal 
E_{PCC}[p]  &= E[p|I(S > k)] w + E[p|I(S \leq k)] (1-w)\\
\eal\eeqa

\noindent and recall that for fixed $k$, $E_{SRS}[p]= E_{PCC}[p]$ when $w = P(S>k)$. For the Binary Entropy criterion defined as $\phi^B(x) = -x \log_2(x) - (1-x) \log_2(1-x)$, $\phi^B(p(S)) = f(S)$ is a function of the scores. For the same fixed $k$, consider the alternative PCC configurations of either $w > P(S>k)$ or $w < P(S>k)$.\\

\noindent \underline{\textit{Case: $w > P(S>k)$}} Using $w > P(S>k)$ implies that $E_{PCC}[p] > E_{SRS}[p]$. Therefore, $\phi^B(p(S)|PCC) > \phi^B(p(S)|SRS))$ in the region of $E[p] < 0.50$ as $H(x)$ is monotone increasing in $x$ when $x < 0.50$. Therefore for rare outcomes using higher stratum weights results in higher sample Binary Entropy.\\

\noindent \underline{\textit{Case: $w < P(S>k)$}} Using $w < P(S>k)$ implies that $E_{PCC}[p] < E_{SRS}[p]$. Therefore, $\phi^B(p(S)|PCC) > \phi^B(p(S)|SRS))$ in the region of $E[p] < 0.50$ as $\phi^B(x)$ is monotone decreasing in $x$ when $x < 0.50$. Therefore for very prevalent outcomes using lower stratum weights results in higher sample Binary Entropy.

\newpage 
\subsection{Additional simulation results for model recalibration}
\label{pcc_sec:appendix_3}

\begin{figure}[!h]
\centering 
\includegraphics[width=6in,height=5in]{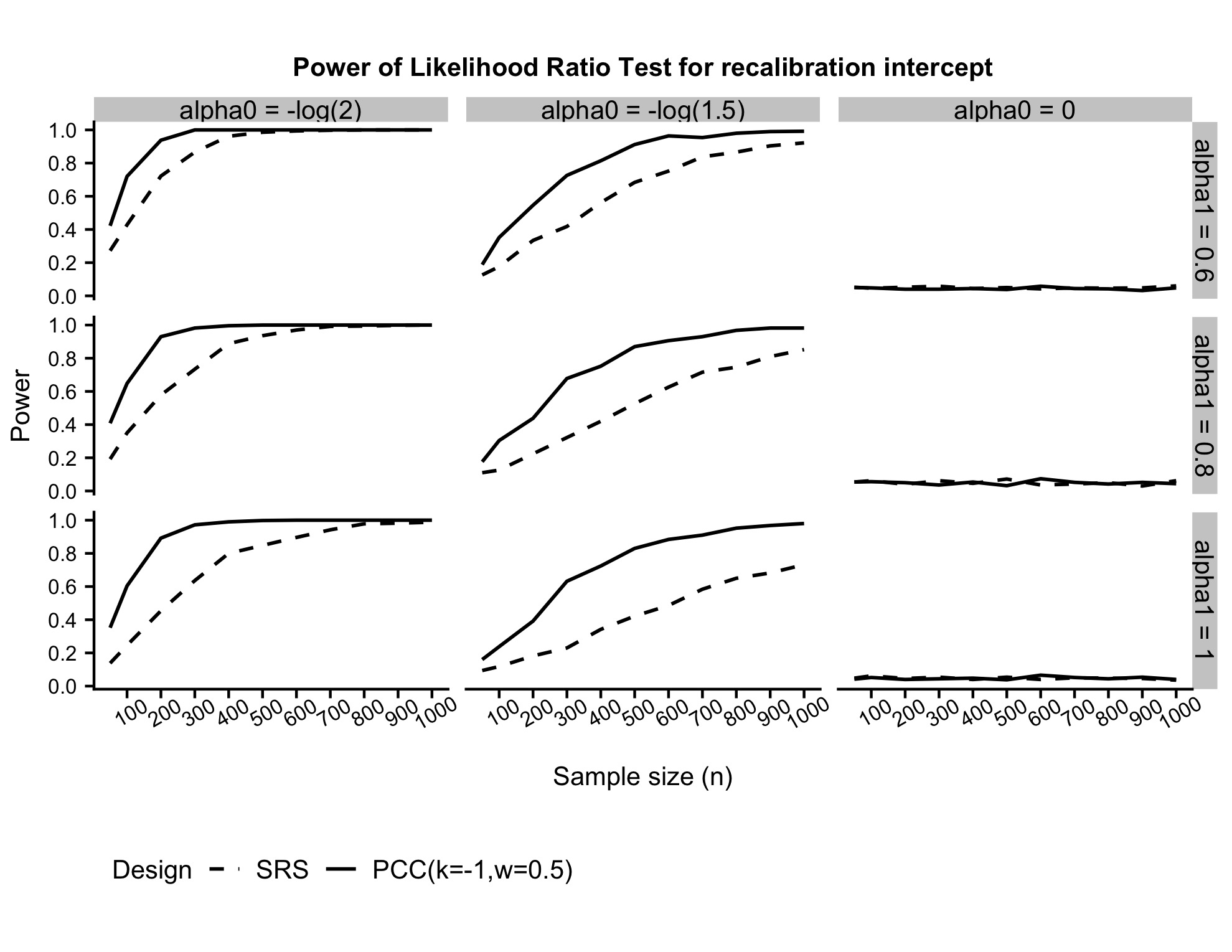}
\end{figure}

\noindent Comments on results:
\bi 
\item The power function under the PCC design is generally higher than under the SRS design.
\item PCC benefit more pronounced for smaller effect size $\alpha_0 = -\log(1.5)$ compared to $-\log(2)$.
\item Both designs provide correct test size.
\ei 

\newpage 
\begin{figure}[!h]
\centering 
\includegraphics[width=6in,height=5in]{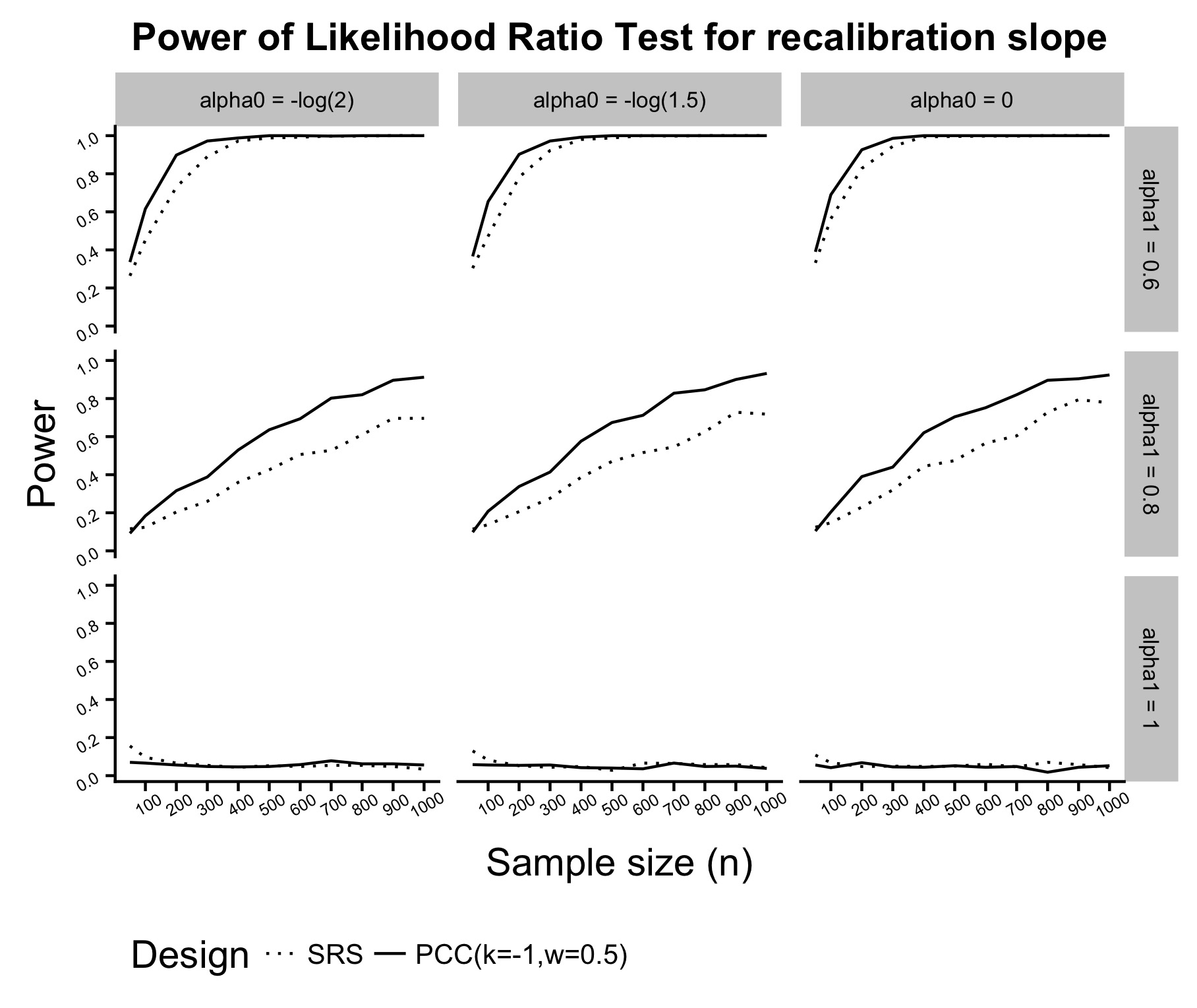}
\end{figure}

\noindent Comments on results:
\bi 
\item The power function under the PCC design is generally higher than under the SRS design.
\item PCC benefit more pronounced for smaller effect size $\alpha_1 = 0.80$ compared to $0.60$.
\item Both designs provide correct test size.
\ei 

\newpage 
\subsection{Details of true model definitions for data application example}
\label{pcc_sec:appendix_4}

\subsubsection*{Source model definition}

\noindent The source model was estimated with the Logistic Lasso procedure, with source model coefficients $\hat{\tilde{\beta}}$ estimated using:

\beqa\bal 
\hat{\tilde{\beta}} &= \underset{\tilde{\beta}}{argmin}
\left\{ \sum\limits_{i=1}^{n} -Y_i(\beta_0 + \tilde{X}^T_i \tilde{\beta} + 
\log(1 + \exp(\beta_0 + \tilde{X}^T_i \tilde{\beta} )) + 
\lambda||\tilde{\beta}||_1 \right\}.
\eal\eeqa 

\noindent The penalization parameter $\lambda$ was selected with 10-fold cross-validation based on the AUC loss function, using value within 1 standard error (\texttt{lambda.1se}), with cross-validation error plot:

\begin{figure}[!h]
\centering 
\includegraphics[width=4.5in,height=2.5in]{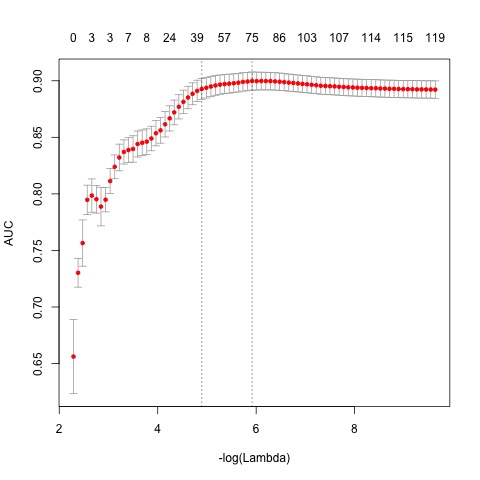}
\end{figure}

\noindent From the selected $38$ features, by thresholding to $0.25$ we obtained a smaller subset of $23$; these coefficients together defined the ``source model'', which were:

\begin{table}[!ht]
\centering
\scriptsize  
\begin{tabular}{lr}
  \hline
Features & Estimate \\ 
  \hline
Intercept & -2.92 \\ 
anterior & 1.55 \\ 
  compress & 1.45 \\ 
  deform & 1.20 \\ 
  endplat & 0.83 \\ 
  not & 0.81 \\ 
  fractur & 0.67 \\ 
  slight & 0.50 \\ 
  none & 0.42 \\ 
  age\_category60+ & 0.41 \\ 
  bodi & 0.40 \\ 
  bone & 0.34 \\ 
  sever & 0.33 \\ 
  desicc & 0.31 \\ 
  later & -0.26 \\ 
  spur & -0.28 \\ 
  evid & -0.29 \\ 
  miner & -0.30 \\ 
  intact & -0.31 \\ 
  sclerosi & -0.32 \\ 
  degen & -0.38 \\ 
  preserv & -0.48 \\ 
  maintain & -0.51 \\ 
  no & -1.09 \\ 
   \hline
\end{tabular}
\end{table}

\subsubsection*{Modification learning model definition}

\noindent The scores were computed as $S_i = \tilde{X}^T_i \hat{\tilde{\beta}}$. Note that the AUC in discriminating cases (Y=1) and controls (Y=0) by using $S$ as the single test was $0.83$. The score distribution by case/control status was:

\begin{figure}[!h]
\centering 
\includegraphics[width=4in,height=2.5in]{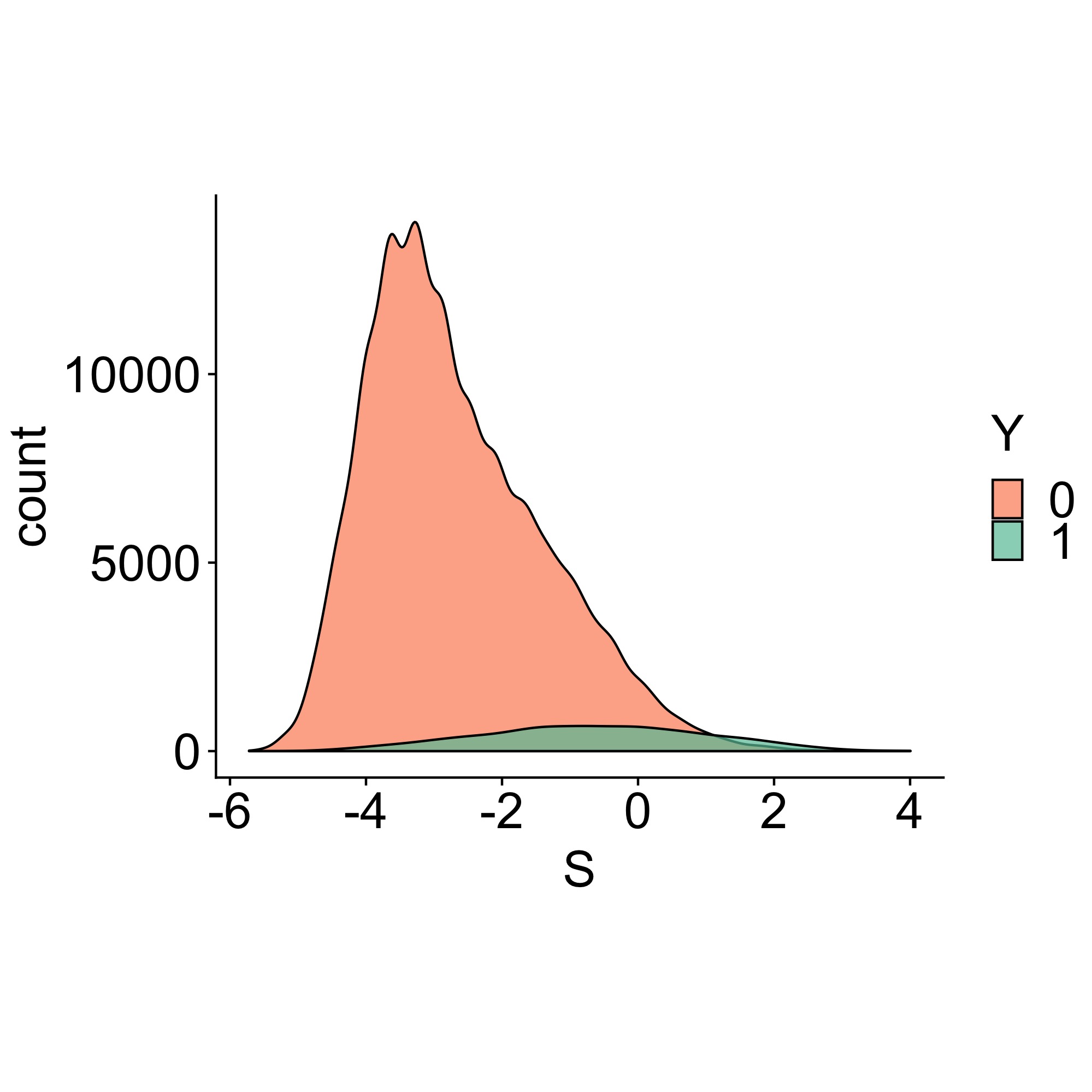}
\end{figure}

\noindent The recalibration parameters were estimated to be $\alpha_0 = -1.00$, $\alpha_1 = 0.89$ using $logit(E[Y_i|S_i]) = \alpha_0 + \alpha_1 S_i$. The revision parameters were estimated with $logit(E[Y_i|S_i,\tilde{X}^T_i]) = \alpha_0 + \alpha_1 S_i + \tilde{X}^T_i \tilde{\gamma}$ with

\beqa\bal 
(\hat{\alpha}_0,\hat{\alpha}_1,\hat{\tilde{\gamma}}) &= \underset{\alpha_0,\alpha_1,\tilde{\gamma}}{argmin}
\left\{ \sum\limits_{i=1}^{n} -Y_i(\alpha_0 + \alpha_1 S_i + \tilde{X}^T_i \tilde{\gamma} + 
\log(1 + \exp(\alpha_0 + \alpha_1 S_i + \tilde{X}^T_i \tilde{\gamma} )) + 
\lambda||\tilde{\gamma}||_1 \right\},
\eal\eeqa 

\noindent From the selected $64$ features, by thresholding to $0.50$ we obtained a smaller subset of $6$; these coefficients together defined the ``target model'', which were:

\begin{table}[!ht]
\centering
\scriptsize  
\begin{tabular}{lr}
  \hline
Feature & Estimate \\  \hline 
 sublux & 1.42 \\ 
deform & 1.15 \\ 
fractur & 0.78 \\ 
scoliosi & 0.78 \\ 
normal & -0.53 \\ 
desicc & -0.57 \\ 
   \hline 
\end{tabular}
\end{table}

\bibliography{master}

\begin{thebibliography}{34}
\providecommand{\natexlab}[1]{#1}
\providecommand{\url}[1]{\texttt{#1}}
\expandafter\ifx\csname urlstyle\endcsname\relax
  \providecommand{\doi}[1]{doi: #1}\else
  \providecommand{\doi}{doi: \begingroup \urlstyle{rm}\Url}\fi

\bibitem[Atkinson(1982)]{atkinson1982developments}
AC~Atkinson.
\newblock Developments in the design of experiments, correspondent paper.
\newblock \emph{International Statistical Review/Revue Internationale de
  Statistique}, pages 161--177, 1982.

\bibitem[Atkinson and Woods(2015)]{atkinson2015designs}
Anthony~C Atkinson and David~C Woods.
\newblock Designs for generalized linear models.
\newblock \emph{Handbook of Design and Analysis of Experiments}, pages
  471--514, 2015.

\bibitem[Batista et~al.(2004)Batista, Prati, and Monard]{batista2004study}
Gustavo~EAPA Batista, Ronaldo~C Prati, and Maria~Carolina Monard.
\newblock A study of the behavior of several methods for balancing machine
  learning training data.
\newblock \emph{ACM Sigkdd Explorations Newsletter}, 6\penalty0 (1):\penalty0
  20--29, 2004.

\bibitem[Chen et~al.(2013)Chen, Carroll, Hinz, Shah, Eyler, Denny, and
  Xu]{chen2013applying}
Yukun Chen, Robert~J Carroll, Eugenia R~McPeek Hinz, Anushi Shah, Anne~E Eyler,
  Joshua~C Denny, and Hua Xu.
\newblock Applying active learning to high-throughput phenotyping algorithms
  for electronic health records data.
\newblock \emph{Journal of the American Medical Informatics Association},
  20\penalty0 (e2):\penalty0 e253--e259, 2013.

\bibitem[Chipman and Welch(1996)]{chipman1996d}
Hugh~A Chipman and William~J Welch.
\newblock D-optimal design for generalized linear models.
\newblock \emph{Unpublished}, 1996.

\bibitem[Cox(1958)]{cox1958two}
David~R Cox.
\newblock Two further applications of a model for binary regression.
\newblock \emph{Biometrika}, 45\penalty0 (3/4):\penalty0 562--565, 1958.

\bibitem[Dalton(2013)]{dalton2013flexible}
Jarrod~E Dalton.
\newblock Flexible recalibration of binary clinical prediction models.
\newblock \emph{Statistics in medicine}, 32\penalty0 (2):\penalty0 282--289,
  2013.

\bibitem[Eriksson et~al.(2000)Eriksson, Johansson, Kettaneh-Wold, Wikstr{\"o}m,
  and Wold]{eriksson2000design}
L~Eriksson, E~Johansson, N~Kettaneh-Wold, C~Wikstr{\"o}m, and S~Wold.
\newblock Design of experiments.
\newblock \emph{Principles and Applications, Learn ways AB, Stockholm}, 2000.

\bibitem[Esteva et~al.(2017)Esteva, Kuprel, Novoa, Ko, Swetter, Blau, and
  Thrun]{esteva2017dermatologist}
Andre Esteva, Brett Kuprel, Roberto~A Novoa, Justin Ko, Susan~M Swetter,
  Helen~M Blau, and Sebastian Thrun.
\newblock Dermatologist-level classification of skin cancer with deep neural
  networks.
\newblock \emph{Nature}, 542\penalty0 (7639):\penalty0 115--118, 2017.

\bibitem[Hosmer et~al.(1997)Hosmer, Hosmer, Le~Cessie, Lemeshow,
  et~al.]{hosmer1997comparison}
David~W Hosmer, Trina Hosmer, Saskia Le~Cessie, Stanley Lemeshow, et~al.
\newblock A comparison of goodness-of-fit tests for the logistic regression
  model.
\newblock \emph{Statistics in medicine}, 16\penalty0 (9):\penalty0 965--980,
  1997.

\bibitem[Jarvik et~al.(2015)Jarvik, Comstock, James, Avins, Bresnahan, Deyo,
  Luetmer, Friedly, Meier, Cherkin, et~al.]{jarvik2015lumbar}
Jeffrey~G Jarvik, Bryan~A Comstock, Kathryn~T James, Andrew~L Avins, Brian~W
  Bresnahan, Richard~A Deyo, Patrick~H Luetmer, Janna~L Friedly, Eric~N Meier,
  Daniel~C Cherkin, et~al.
\newblock Lumbar imaging with reporting of epidemiology (lire)—protocol for a
  pragmatic cluster randomized trial.
\newblock \emph{Contemporary clinical trials}, 45:\penalty0 157--163, 2015.

\bibitem[Khuri et~al.(2006)Khuri, Mukherjee, Sinha, and Ghosh]{khuri2006design}
Andr{\'e}~I Khuri, Bhramar Mukherjee, Bikas~K Sinha, and Malay Ghosh.
\newblock Design issues for generalized linear models: A review.
\newblock \emph{Statistical Science}, pages 376--399, 2006.

\bibitem[Kiefer and Wolfowitz(1959)]{kiefer1959optimum}
Jack Kiefer and Jacob Wolfowitz.
\newblock Optimum designs in regression problems.
\newblock \emph{The Annals of Mathematical Statistics}, pages 271--294, 1959.

\bibitem[Little and Rubin(2014)]{little2014statistical}
Roderick~JA Little and Donald~B Rubin.
\newblock \emph{Statistical analysis with missing data}.
\newblock John Wiley \& Sons, 2014.

\bibitem[MacKay(2003)]{mackay2003information}
David~JC MacKay.
\newblock \emph{Information theory, inference and learning algorithms}.
\newblock Cambridge university press, 2003.

\bibitem[Moons et~al.(2004)Moons, Donders, Steyerberg, and
  Harrell]{moons2004penalized}
KGM Moons, A~Rogier~T Donders, EW~Steyerberg, and FE~Harrell.
\newblock Penalized maximum likelihood estimation to directly adjust diagnostic
  and prognostic prediction models for overoptimism: a clinical example.
\newblock \emph{Journal of clinical epidemiology}, 57\penalty0 (12):\penalty0
  1262--1270, 2004.

\bibitem[Nguyen and Patrick(2014)]{nguyen2014supervised}
Dung~HM Nguyen and Jon~D Patrick.
\newblock Supervised machine learning and active learning in classification of
  radiology reports.
\newblock \emph{Journal of the American Medical Informatics Association},
  21\penalty0 (5):\penalty0 893--901, 2014.

\bibitem[Pan et~al.(2010)Pan, Yang, et~al.]{pan2010survey}
Sinno~Jialin Pan, Qiang Yang, et~al.
\newblock A survey on transfer learning.
\newblock \emph{IEEE Transactions on knowledge and data engineering},
  22\penalty0 (10):\penalty0 1345--1359, 2010.

\bibitem[Pukelsheim(1993)]{pukelsheim1993optimal}
Friedrich Pukelsheim.
\newblock \emph{Optimal design of experiments}, volume~50.
\newblock siam, 1993.

\bibitem[Schein and Ungar(2007)]{schein2007active}
Andrew~I Schein and Lyle~H Ungar.
\newblock Active learning for logistic regression: an evaluation.
\newblock \emph{Machine Learning}, 68\penalty0 (3):\penalty0 235--265, 2007.

\bibitem[Settles(2012)]{settles2012active}
Burr Settles.
\newblock Active learning.
\newblock \emph{Synthesis Lectures on Artificial Intelligence and Machine
  Learning}, 6\penalty0 (1):\penalty0 1--114, 2012.

\bibitem[Steyerberg(2008)]{steyerberg2008clinical}
Ewout~W Steyerberg.
\newblock \emph{Clinical prediction models: a practical approach to
  development, validation, and updating}.
\newblock Springer Science \& Business Media, 2008.

\bibitem[Steyerberg and Vergouwe(2014)]{steyerberg2014towards}
Ewout~W Steyerberg and Yvonne Vergouwe.
\newblock Towards better clinical prediction models: seven steps for
  development and an abcd for validation.
\newblock \emph{European heart journal}, 35\penalty0 (29):\penalty0 1925--1931,
  2014.

\bibitem[Steyerberg et~al.(2004)Steyerberg, Borsboom, van Houwelingen,
  Eijkemans, and Habbema]{steyerberg2004validation}
Ewout~W Steyerberg, Gerard~JJM Borsboom, Hans~C van Houwelingen, Marinus~JC
  Eijkemans, and J~Dik~F Habbema.
\newblock Validation and updating of predictive logistic regression models: a
  study on sample size and shrinkage.
\newblock \emph{Statistics in medicine}, 23\penalty0 (16):\penalty0 2567--2586,
  2004.

\bibitem[Stukel(1988)]{stukel1988generalized}
Th{\'e}r{\`e}se~A Stukel.
\newblock Generalized logistic models.
\newblock \emph{Journal of the American Statistical Association}, 83\penalty0
  (402):\penalty0 426--431, 1988.

\bibitem[Tan et~al.(2018)Tan, Hassanpour, Heagerty, Rundell, Suri, Huhdanpaa,
  James, Carrell, Langlotz, Organ, et~al.]{tan2018comparison}
W~Katherine Tan, Saeed Hassanpour, Patrick~J Heagerty, Sean~D Rundell, Pradeep
  Suri, Hannu~T Huhdanpaa, Kathryn James, David~S Carrell, Curtis~P Langlotz,
  Nancy~L Organ, et~al.
\newblock Comparison of natural language processing rules-based and
  machine-learning systems to identify lumbar spine imaging findings related to
  low back pain.
\newblock \emph{Academic radiology}, 2018.

\bibitem[Tibshirani(1996)]{tibshirani1996regression}
Robert Tibshirani.
\newblock Regression shrinkage and selection via the lasso.
\newblock \emph{Journal of the Royal Statistical Society. Series B
  (Methodological)}, pages 267--288, 1996.

\bibitem[Tomanek and Olsson(2009)]{tomanek2009web}
Katrin Tomanek and Fredrik Olsson.
\newblock A web survey on the use of active learning to support annotation of
  text data.
\newblock In \emph{Proceedings of the NAACL HLT 2009 workshop on active
  learning for natural language processing}, pages 45--48. Association for
  Computational Linguistics, 2009.

\bibitem[Vergouwe et~al.(2005)Vergouwe, Steyerberg, Eijkemans, and
  Habbema]{vergouwe2005substantial}
Yvonne Vergouwe, Ewout~W Steyerberg, Marinus~JC Eijkemans, and J~Dik~F Habbema.
\newblock Substantial effective sample sizes were required for external
  validation studies of predictive logistic regression models.
\newblock \emph{Journal of clinical epidemiology}, 58\penalty0 (5):\penalty0
  475--483, 2005.

\bibitem[Wald(1943)]{wald1943tests}
Abraham Wald.
\newblock Tests of statistical hypotheses concerning several parameters when
  the number of observations is large.
\newblock \emph{Transactions of the American Mathematical society}, 54\penalty0
  (3):\penalty0 426--482, 1943.

\bibitem[Weiss and Provost(2001)]{weiss2001effect}
Gary~M Weiss and Foster Provost.
\newblock The effect of class distribution on classifier learning: an empirical
  study.
\newblock 2001.

\bibitem[Xue and Hall(2015)]{xue2015does}
Jing-Hao Xue and Peter Hall.
\newblock Why does rebalancing class-unbalanced data improve auc for linear
  discriminant analysis?
\newblock \emph{IEEE transactions on pattern analysis and machine
  intelligence}, 37\penalty0 (5):\penalty0 1109--1112, 2015.

\bibitem[Yang and Loog(2018)]{yang2018benchmark}
Yazhou Yang and Marco Loog.
\newblock A benchmark and comparison of active learning for logistic
  regression.
\newblock \emph{Pattern Recognition}, 83:\penalty0 401--415, 2018.

\bibitem[Zadrozny(2004)]{zadrozny2004learning}
Bianca Zadrozny.
\newblock Learning and evaluating classifiers under sample selection bias.
\newblock In \emph{Proceedings of the twenty-first international conference on
  Machine learning}, page 114. ACM, 2004.

\end{thebibliography}

\end{document}